\documentclass[fleqn,usenatbib]{mnras}

\usepackage{newtxtext,newtxmath}

\usepackage[utf8]{inputenc}
\usepackage[T1]{fontenc}
\usepackage[dvipsnames]{xcolor}

\usepackage{graphicx}	
\usepackage{amsmath}	
\usepackage{float}

\usepackage{xspace}

\usepackage[normalem]{ulem}
\usepackage{physics}
\usepackage{siunitx}
\newcommand{\galsim}{\texttt{GalSim}\xspace}
\newcommand{\scarlet}{\texttt{\scshape scarlet}\xspace}
\usepackage{fontawesome}

\usepackage{graphicx}
\usepackage{color}
\usepackage{tikz}
\usepackage{pgf-umlsd}
\usepackage{ifthen}

\usepackage{cleveref}

\newcommand{\github}{\href{https://github.com/LSSTDESC/DeblenderVAE}{\faGithub}}
\crefformat{footnote}{#2\footnotemark[#1]#3}
\newcommand*{\eg}{e.g.\@\xspace}
\newcommand*{\ie}{i.e.\@\xspace}

\title[Deblending galaxies with VAE]{Deblending galaxies with Variational Autoencoders: a joint multi-band, multi-instrument approach}

\author[Arcelin et al.]{
Bastien Arcelin,$^{1}$\thanks{E-mail: arcelin@apc.in2p3.fr (APC)}
Cyrille Doux,$^{1,2}$
Eric Aubourg$^{1}$
and C\'ecile Roucelle$^{1}$
\newauthor
(The LSST Dark Energy Science Collaboration)
\\
$^{1}$Université de Paris, CNRS, Astroparticule et Cosmologie, F-75013 Paris, France
\\
$^{2}$Center for Particle Cosmology, Department of Physics and Astronomy, University of Pennsylvania, 209 South 33rd Street, Philadelphia, PA 19104, USA\\
}

\date{Accepted XXX. Received YYY; in original form ZZZ}

\pubyear{2020}

\begin{document}
\label{firstpage}
\pagerange{\pageref{firstpage}--\pageref{lastpage}}
\maketitle

\begin{abstract}
Blending of galaxies has a major contribution in the systematic error budget of weak lensing studies, affecting photometric and shape measurements, particularly for ground-based, deep, photometric galaxy surveys, such as the Rubin Observatory Legacy Survey of Space and Time (LSST). Existing deblenders mostly rely on analytic modelling of galaxy profiles and suffer from the lack of flexible yet accurate models. We propose to use generative models based on deep neural networks, namely variational autoencoders (VAE), to learn probabilistic models directly from data. We train a VAE on images of centred, isolated galaxies, which we reuse, as a prior, in a second VAE-like neural network in charge of deblending galaxies. We train our networks on simulated images including six LSST bandpass filters and the visible and near-infrared bands of the Euclid satellite, as our method naturally generalises to multiple bands and can incorporate data from multiple instruments. We obtain median reconstruction errors on ellipticities and $r$-band magnitude between $\pm\num{0.01}$ and $\pm\num{0.05}$ respectively in most cases, and ellipticity multiplicative bias of 1.6\% for blended objects in the optimal configuration. We also study the impact of decentring and prove the method to be robust. This method only requires the approximate centre of each target galaxy, but no assumptions about the number of surrounding objects, pointing to an iterative detection/deblending procedure we leave for future work. Finally, we discuss future challenges about training on real data and obtain encouraging results when applying \textit{transfer learning}. Our code is publicly available on GitHub \github.
\end{abstract}




\section{Introduction}

Upcoming galaxy surveys, such as the Legacy Survey of Space and Time \citep[LSST,][]{2009arXiv0912.0201L} conducted at the future Vera C. Rubin Observatory, ESA's Euclid satellite \citep{2010arXiv1001.0061R} and the Wide-Field Infrared Survey Telescope \citep[WFIRST,][]{2013arXiv1305.5422S} will produce an unprecedented amount of observations and astrophysical data and will be a major step forward in physical cosmology.
They will be used to study dark energy with various precision cosmological probes \citep[see, for instance,][]{2013FrPhy...8..828L} and in particular cosmic shear, which measures the coherent distortion of background galaxies by foreground matter through weak gravitational lensing. 
It has already become a powerful observable to constrain cosmology \citep[see][for earlier studies]{2002PhRvD..65b3003H,2002PhRvD..65f3001H} and is now systematically used in precursor surveys such as in CFHTLenS \citep{2017MNRAS.465.2033J}, DES \citep{2018arXiv181002499D} or HSC \citep{2019PASJ...71...43H}.

With increased depth, larger sky coverage and higher resolution (for space-based experiments), the next generation of surveys is expected to bring this probe to unprecedented levels of precision. However, the efficient treatment and reduction of the systematic effects that will arise as the main source of error \citep[see][for a detailed review]{Mandelbaum2017} will be key for their scientific return.
With wide and deep surveys, one will detect many more faint galaxies and low surface brightness features for extended objects than with the previous-generation instruments. 
As a consequence, the apparent blending of galaxies with other astrophysical sources (mainly other galaxies and stars) aligned along the line of sight will become prevalent. This effect will even be reinforced for the LSST survey in comparison to space-based surveys because of the degradation of the point-spread function (PSF) by the atmosphere. \citet{Bosch2017} estimate that 58\% of objects detected in the HSC wide survey are blended. As LSST expects to reach $i\sim27$ after 10 years of operations (compared to $i\sim26$ for HSC) an even greater fraction of blended objects is to be expected. In this paper, we focus on the specific problem of blending between multiple galaxies, as extended objects require a different processing than point sources such as stars (and other sources like asteroid trails).

In the prospect of weak lensing analyses, blending affects measurements of shapes and fluxes of blended galaxies and therefore impacts cosmic shear measurements by either degrading statistics or introducing biases, as the vast majority of measurement algorithms are designed for isolated objects.
To exploit the new generation of surveys to their full statistical strength without impacting the systematic error budget, it is therefore of primary importance to possess efficient methods to perform measurements on blended objects. 
A path to explore is to \textit{deblend} galaxies, \ie to reconstruct from a detected blended scene the images of each isolated galaxy and run existing measurement pipelines on those images. 
Current deblending methods such as \texttt{SExtractor} \citep{1996A&AS..117..393B} and the deblender of the SDSS pipeline (Lupton, in prep.) are based on known properties of galaxies (luminosity, central symmetry, etc.) and most of them use only one bandpass filter. More recently, \citet*{2016A&A...589A...2J} with \texttt{MuSCADeT} and \cite{scarlet} with \scarlet, have proposed methods using several bandpass filters but they are still based on assumptions about galaxy properties. For instance, \texttt{MuSCADeT} assumes sparsity within a template library while \scarlet can assume monotonicity and/or central symmetry of the light profile). On the other hand,  machine-learning techniques based on neural networks can be used to build a model of galaxy images by learning abstract features from a sample of images, without making any particular assumption, and naturally generalise to multi-channel images. This allows ones to build a tool suitable for ingesting multiband images from a single survey and/or for combining multiple observations for the same object from different instruments at the frame level. Additionally, neural networks are execute quickly once trained, a key advantage to process the volume of data from upcoming surveys. As a consequence, they appear to be a sensible tools to tackle the deblending challenge and particularly relevant in our context.  \citet{2018arXiv181010098R} indeed recently showed that it is possible to deblend low-redshift galaxies using a branched Generative Adversarial Network (GAN) on RGB images (i.e. three-bands images) and \citet{2020MNRAS4912481B} used a U-Net architecture \citep{2015arXiv150504597R} to perform segmentation on single-band images and recover the photometry of high-redshift blended galaxies. 

In this paper, we propose another method based on
a different kind of generative neural network called
\textit{variational autoencoders} \citep[VAE, proposed by][]{2013arXiv1312.6114K} that have become extremely popular among generative models in the past years. In practice, VAE are made of two components: a generative model, mapping latent variables to a likelihood in data space, and a recognition model, learning a posterior distribution of latent variables conditioned on data.
\citet{2016arXiv160905796R} showed that VAE can be used as deep (conditional) generative models in producing isolated galaxy images and that they produce more consistent results than GAN. Additionally, VAE have the advantage of providing an approximate likelihood of image samples and predicting posterior distributions.
We propose a method based on two VAE-like neural networks to address the deblending challenge and we perform an initial study of performance on simulated data.\\
More precisely, the method is developed in two steps: 
\begin{enumerate}
    \item First, we train a VAE to learn the features of isolated galaxy images and their representation in a latent space. The aim of this network is to build a generative model for isolated galaxy images directly from the (here simulated) data. 
    The images are generated with LSST and Euclid bandpass filters with fixed PSF (see \cref{sec:data_sample}).
    \item Second, we use this trained VAE to create a second neural network, with a similar architecture, in charge of performing the deblending task. The generative component of the VAE is used with fixed parameters (\ie non-trainable) and the recognition model is retrained, though with noisy images of blended galaxies as input. This method allows us to use the generative model previously built as a prior for isolated galaxies images and consequently to perform deblending. The artificially blended images are also generated with LSST and Euclid bandpass filters. We will evaluate the performance of the method in terms of errors introduced on intrinsic ellipticities (deconvolved from the PSF) and fluxes (or magnitudes) per band, which are relevant parameters for weak lensing studies.
\end{enumerate}

Using these two networks, we evaluate the performance of a multi-band analysis using the six LSST bandpass filters jointly in the construction of a prior for single galaxy images and in deblending galaxies by considering shape and flux reconstruction of deblended objects.
Moreover \citet*{2019arXiv190108586S} showed that combining a ground-based and space-based experiment like LSST and Euclid in a joint-pixel analysis can significantly improve shape measurements compared to the independent analysis of LSST. We therefore quantify the extent to which LSST can benefit from Euclid for the sake of deblending galaxies.

Moreover, we need to provide the deblender network with information to determine which galaxy is to be deblended in a given scene, which is done by centring the image stamp on that particular galaxy. The analysis presented in this paper is realised on images where, in the fiducial case, the target galaxy is perfectly centred on isolated galaxy images as well as on blended images. However this ideal case is very unlikely in reality as noise, pixelisation or geometric properties of the galaxy (\eg asymmetry) and its neighbours, for example, impact the target galaxy centring even in the case of an extremely accurate detection algorithm. Consequently, we also estimate the impact of decentring with two other cases: the first one supposing a perfect detection algorithm and taking into account only noise and pixelisation, and the second one using a basic detection algorithm to centre our stamps, as could be performed on real data.

Note that only galaxy-galaxy separation is studied in this paper as star-galaxy separation is considered in the literature \citep[\texttt{SCARLET}]{scarlet} to be less difficult (shapes and light profiles of stars being arguably simpler). Nevertheless, as our method focuses on recovering the brightest, centred galaxy and removing all other components, we expect it should generalise to star-galaxy separation, provided stars are included in the training sample.

The paper is organised as follows: \cref{sec:data_sample} presents the generation of the training and test samples from the HST COSMOS catalogue\footnote{Available at \url{https://zenodo.org/record/3242143}.} \citep{mandelbaum_rachel_2012_3242143}  using \galsim \citep{2015A&C....10..121R}. \Cref{sec:method} presents our networks and the specific architectures we used in this study. \Cref{sec:results} details results obtained and the comparison of multi-bands and multi-survey analyses as well as the impact of decentring on our method. Finally, we discuss our results in \cref{sec:discussion} and present our conclusions in \cref{sec:conclusion}

\section{Simulated images}
\label{sec:data_sample}

In this section, we describe the generation of simulated images for isolated and blended galaxies using \galsim\footnote{\url{https://github.com/GalSim-developers/GalSim}} \citep{2015A&C....10..121R}. These simulated images are based on parametric models fitted to real galaxies from the HST COSMOS catalogue containing \num{81500} images with F814W<\num{25.2}. These fits were realised for the third Gravitational Lensing Accuracy Testing (GREAT3) Challenge \citep[see Appendix E.2]{2014ApJS..212....5M}. Each COSMOS galaxy is fitted twice --once with a Sérsic profile (with index, $n$, free), and once with a de Vaucouleurs ($n=4$) bulge profile plus an exponential ($n=1$) disk profile-- and the best of the two fits is kept. In addition to central position and ellipticities, the free parameters for Sérsic profiles are effective radius $R_{\rm eff}$, intensity $I_{1/2}$ and index $n$. \galsim can generate multi-band images with PSF models and pixel noise chosen to simulate LSST-like and Euclid-like images, based on parametric profiles fitted to observations.

\subsection{Datasets}
Two sets of images have been generated for the training phase: a training (\num{100000} images) and a validation sample (\num{20000} images). A third independent test sample (\num{10000} images) has been generated with 5000 COSMOS galaxies that were not used for the training and validation samples. The performance of our networks is measured on this test sample. For each sample, we simulate both noisy and noiseless images, the former being used as inputs and the latter as targets (i.e. images to reproduce, used for computation in the loss function) of the neural networks.

\subsubsection{Point-spread functions and noise}
\label{sec:PSF_noise}
For LSST bandpass filters, we use a fixed Kolmogorov point-spread function (PSF), implemented in \galsim, using a full width at half maximum (FWHM) of \SI{0.65}{\arcsecond}, which is the median value of the expected FWHM distribution, as shown in \citet{2008arXiv0805.2366I}. This is an approximation as our LSST images are a stack of a large number of exposures in each filter. In practice, the PSF is varying, due to changing weather conditions on each frame, and potentially anisotropic. Therefore, the effective PSF on a stacked image would not be exactly the median of the distribution and would vary from one stacked image to another. We reserve this question for future work but note that, for practical applications, two options exist: images could be deconvolved from the measured PSF and reconvolved with the fixed PSF we use, before being passed into the deblender network or the training sample could be augmented with multiple realisations of the PSF for each scene.

For Euclid instruments, the PSF is much more stable as it is a space-based experiment and it can be considered fixed to a very good approximation. The PSF is applied to the filters using the Moffat model from \galsim with an FWHM of, respectively, \SI{0.22}{\arcsecond} and \SI{0.18}{\arcsecond} \citep{2010arXiv1001.0061R} for the NISP and VIS instruments.

For each of the ten bandpass filters, Poisson noise is added, taking into account the background sky level\footnote{Data is available at \url{https://smtn-002.lsst.io/} for LSST.} (appropriately scaled to account for exposure time and instrument properties). The expected value of the sky background is then subtracted from the image in order to centre the noise distribution around 0. We note that noise in the actual Euclid and LSST stacks may significantly be correlated from pixel to pixel due to image resampling, an effect that we do not consider in this work.

\subsubsection{Training set 1: isolated galaxies}
\label{sec:isolated_galaxy_generation}
\begin{figure}
	\includegraphics[trim={0.5cm 1cm 0.5cm 0cm}, width=\columnwidth]{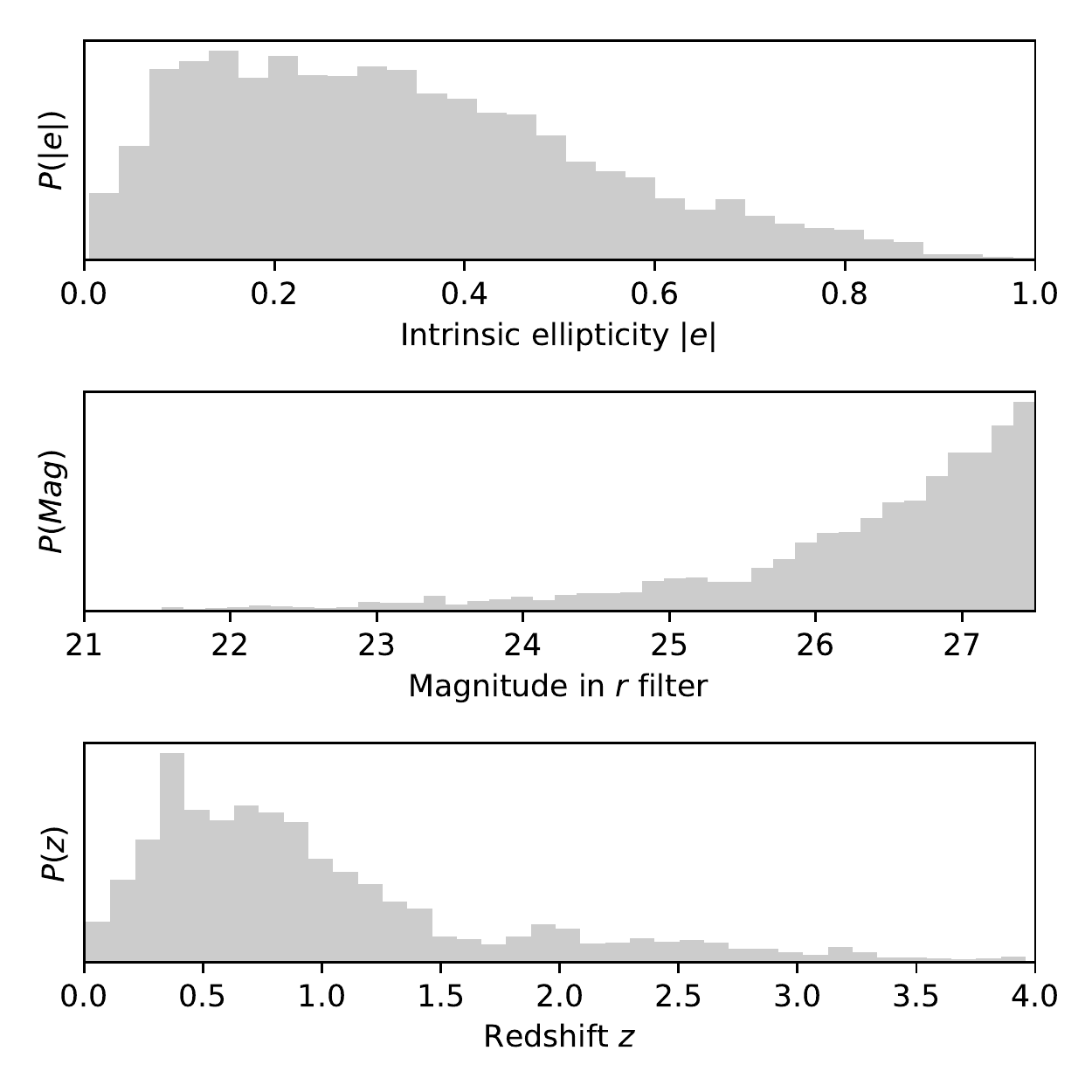}
    \caption{Distributions of observed ellipiticity, $r$-band magnitude (a cut is applied at 27.5) and redshift for galaxies in the test sample.}
    \label{fig:param_distrib}
\end{figure}

Using the \texttt{galsim.COSMOSCatalog} class, we generate images of isolated galaxies from the parametric models that have been fitted to the COSMOS catalogue. During image generation, the dataset of \num{81500} galaxies is randomised by applying rotations before rendering and drawing Poisson noise realisations (see \cref{sec:PSF_noise}). In addition, during training and validation runs, we randomly flip images horizontally and vertically and rotate them by \SI{90}{\degree} on the fly since these operations require no resampling.
The spectral energy distributions (SED), profiles and shapes of each galaxy are directly extracted from the catalogue (see \cref{fig:param_distrib}).

Images are generated for the six LSST bandpass filters ($u$, $g$, $r$, $i$, $z$ and $y$) as defined in \galsim and for the four Euclid bandpass filters\footnote{Available at \url{http://svo2.cab.inta-csic.es/svo/theory/fps/index.php?mode=browse&gname=Euclid}.} (three for the Near Infrared Spectrometer and Photometer, NISP, one for the visible instrument, VIS). Note that since we use parametric profiles, channels only differ by flux and noise realisations, \ie we do not simulate colour gradients in this work. Image size is fixed to 64x64 pixels for each instrument which corresponds to, respectively, \SIlist{19.2;12.8;6.4}{\arcsecond}-wide images for the NIR filters, for the LSST filters and for the VIS instrument, given their resolutions of 0.3, 0.2 and 0.1 arcsecond per pixel. This fixed stamp size was chosen to work with simple network architectures, although in future work, several stamp sizes might be used for each instrument.

For each galaxy, we consider a stack of four \SI{450}{\s} exposures for Euclid \citep[see \textit{Images Simulations} in][]{2010arXiv1001.0061R}, and a stack of 824 15-second exposures for LSST, split unevenly between bands, though constant between samples, to match the projected full LSST survey mean exposure \citep[specifically, we use respectively 56, 80, 184, 184, 160 and 160 exposures for the $ugrizy$ bands as in][table 1]{2008arXiv0805.2366I}. The flux of each galaxy is rescaled from HST COSMOS observations to account for exposure time, number of exposures, pixel size and primary mirror area of each instrument. Finally, we apply a cut in the LSST $r$-band magnitude of 27.5, at the LSST fiducial depth\footnote{See, for instance, \url{https://www.lsst.org/scientists/keynumbers}}.

\subsubsection{Training set 2: blended galaxies}
\label{deblender_training}
The deblender network is trained and validated on samples of artificially blended galaxies obtained by adding images of isolated galaxies, produced as described above (thus neglecting galaxy opacity, as done in most of the deblending literature at this point).
The number of galaxies on each image is randomly chosen between 1 and 4 with uniform probabilities.
From these galaxies, the one with the smallest magnitude is centred in the blended image and is the network's target for reconstruction (see \cref{sec:decentring_data} for more details).
As for images of single galaxies, a cut of $r<27.5$ is applied for all galaxies added to the blend. 
Non-centred galaxies are randomly positioned in an annulus around the centred galaxy, with inner radius equal to half of the expected median LSST PSF full-width-at-half-maximum (that is, \SI{0.325}{\arcsecond}) and outer radius equal to \SI{2}{\arcsecond}. This choice allows us to obtain highly blended galaxies while avoiding cases where non-centred galaxies completely overlap the centred one.

Note that for these samples, SNR and magnitude distributions are slightly shifted to higher SNR/smaller magnitudes since we always choose the smallest magnitude galaxies as targets, but the ranges of SNR and magnitudes remain unchanged.

Finally, as for isolated galaxy images generation, we apply randomisation to the dataset before rendering, and data augmentation (random flips and \SI{90}{\degree} rotations) during training, the latter being particularly relevant for blended scenes.
 
\subsubsection{Decentring}
\label{sec:decentring_data}

In the fiducial case, considered as reference in this paper, target galaxies are perfectly centred on the stamp (at the intersection of the four central pixels).
In order to evaluate the impact of decentring, we created two additional configurations, which are defined as follows.
\begin{enumerate}
    \item The first one considers that target galaxies are identified as the ones with smallest magnitudes. Their centres are shifted uniformly around the stamp centre within a square of size \SI{0.2}{\arcsecond}, \ie the size of LSST's pixels. The shift is applied for isolated galaxies images as well as for blended galaxies images. This configuration simulates a perfectly accurate peak detection algorithm where decentring is only due to pixelisation.
    \item In the second configuration, we start by producing blended images as in the first configuration, albeit with 128x128 pixels, and subsequently process them with the peak detection algorithm from the  \texttt{photutils}\footnote{\url{https://photutils.readthedocs.io/en/stable/}} library. We identify the galaxy to be deblended as the galaxy with the brightest detected centroid in the $r$-bandpass filter. We then determine the closest pixel intersection and crop 64x64 images centred on this point.
    For isolated galaxy images, the galaxies are shifted from the stamp centre a first time, as in the first configuration, for pixelisation. They are then shifted a second time with a value sampled from a distribution fitting centring errors in the blended sample \citep[in practice, a beta prime distribution][]{johnson1995continuous}, before images are finally drawn. This case is meant to mimic more closely a detection-deblending pipeline that could be applied to real data where we do not know \textit{a priori} which galaxy is the brightest within a blended scene. It is fairly conservative in terms of decentring as we used a basic peak detection algorithm on the $r$ bandpass filter only. Note, however, that this procedure slightly redefines the objective of the deblender network as we are now targeting the galaxy with the brightest detected centroid which is different from the one with the smallest magnitude in about 20\% of cases in our samples. We discuss these points in greater detail in \cref{sec:discussion_decentring}.
\end{enumerate} 

\subsection{Blendedness metrics}
\label{sec:blendedness_metrics}

We define several metrics to characterise how much galaxies are blended, all of which are computed in the LSST $r$-band (on which we will perform ellipticity and magnitude measurements) on PSF-convolved images.
\begin{enumerate}
    \item First the blending rate of the centred galaxy with its closest neighbour is defined as
    \begin{equation}
        B_{\rm closest} = \frac{\expval{I_{\rm centred}, I_{\rm closest}}}{\sqrt{\expval{I_{\rm centred}, I_{\rm centred}} \expval{I_{\rm closest}, I_{\rm closest}}}} 
        \label{eq:blendrate_closest}
    \end{equation}
    where the dot product is defined as
    \begin{equation}
        \expval{I, I^\prime} = \sum_{p \in \{\rm pixels\}} I_p I_p^\prime.
    \end{equation}
    $I_{\rm centred}$ represents the noiseless image of the centred galaxy and $I_{\rm closest}$ represents the noiseless image of the closest (centre-wise) blended galaxy.
    \item Second, a \textit{total} blending rate is defined using the definition proposed in \citet{scarlet} as
    \begin{equation}
        B_{\rm tot} = 1 - \frac{\expval{I_{\rm centred}, I_{\rm centred}}}{\expval{I_{\rm centred}, I_{\rm total}}}.
        \label{eq:blendrate_total}
    \end{equation}
\end{enumerate}

These quantities characterise the blendedness of the centred galaxy within a blended scene and have values between 0 and 1. A blend rate close to 0 indicates almost no blending and a blend rate close to 1 indicates that the (closest) neighbouring galaxies and the centred one overlap almost completely. The shifts applied to neighbouring galaxies (see \cref{deblender_training}) was purposefully chosen to obtain a significant amount of highly blended scenes (under those metrics). Note that the hypotheses for images generation are not meant to reproduce the distribution of real blends but rather to permit us to train and test our network on relevant blended scenes, albeit in a simplified framework (fixed PSF, simulated parametric images).

\subsection{Image preprocessing}
\label{sec:batch_generation}
The networks are fed with batches of images randomly chosen from the training sample, that are also randomly flipped and rotated by \SI{90}{\degree} on the fly, as explained in \cref{sec:isolated_galaxy_generation}. Both noisy and noiseless images were preprocessed by applying, in each band $b$, the following normalisation,
\begin{equation}
    \centering
    x_{b} = \tanh(\sinh^{-1}\qty(\beta\frac{x_{{\rm raw},b}}{\expval{\max(x_{{\rm raw},b})}_b})),
    \label{eq:processing}
\end{equation}
where $\expval{\max(x_{{\rm raw},b})}_b$ is the mean of the distribution of maximum pixel values in the $b$ band of input images (therefore a constant defined for each bin).
$\beta$ is an arbitrary constant set to 2.5 for our study allowing our training sample to have maxima well-distributed in the range [0,1].
This processing is necessary because of the large dynamic range of astronomical images, to ensure that images of bright galaxies do not cause numerical instability during training. Even though this normalisation constrains the choice of likelihood distribution (see \cref{sec:architecture_VAE}), we have experimented with non-normalised images and more general distributions and ran into numerical issues (see \cref{sec:towards_bayesian}).

\section{Method}
\label{sec:method}

Two different neural networks are used in this study. The first one is a denoising variational autoencoder \citep[VAE,][]{2013arXiv1312.6114K} which aims to reproduce accurately noiseless images of isolated galaxies from noisy images. The second one is a deblender, with a similar architecture, which aims to reproduce accurately the centred galaxy from a noisy image of blended galaxies.

\begin{figure*}
	\includegraphics[width=\linewidth]{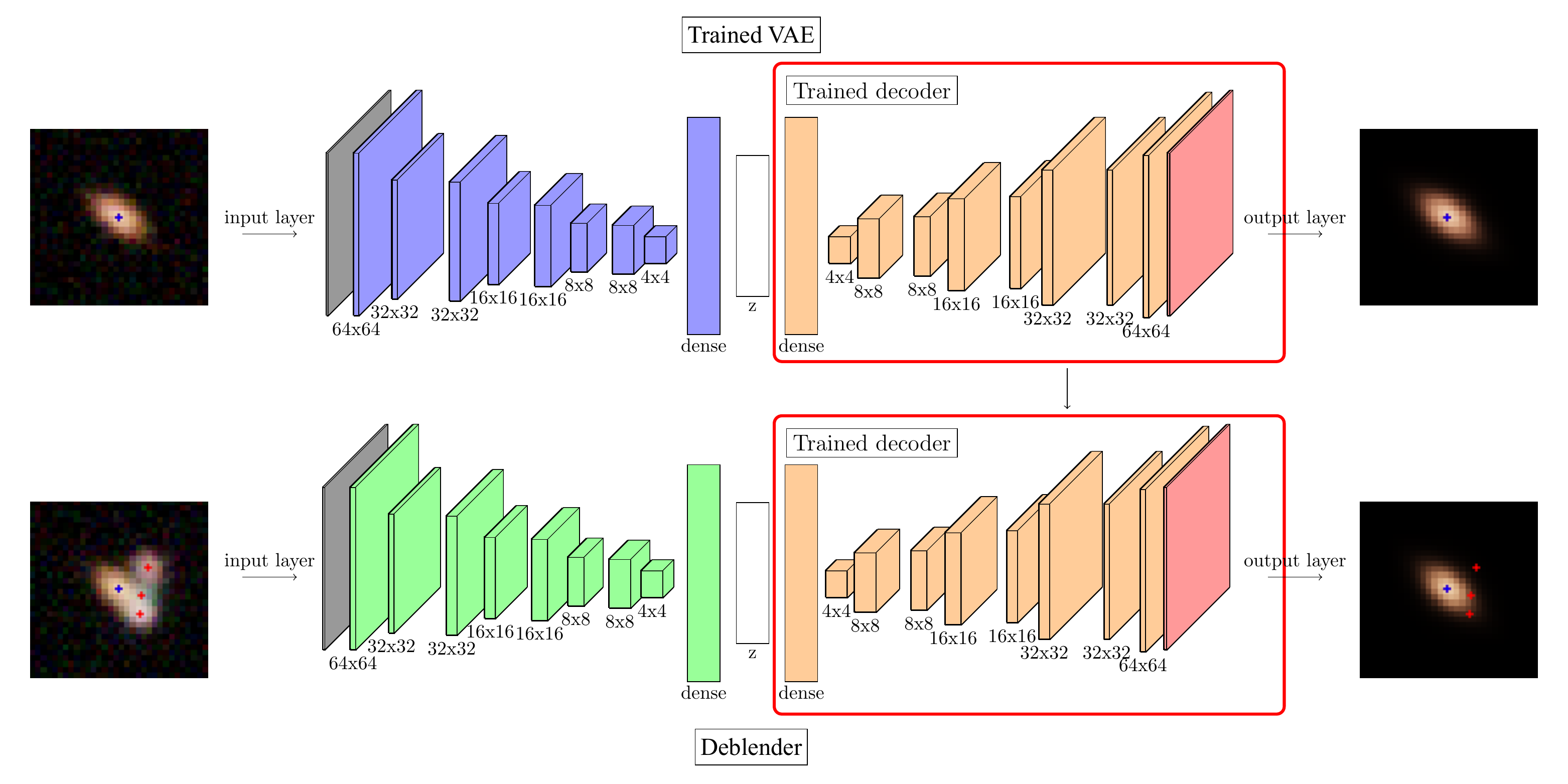}
    \caption{Architectures of the VAE and the deblender. The weights of the VAE's trained decoder are loaded and fixed in the deblender network before training.}
    \label{fig:VAE_deblender_architecture}
\end{figure*}

\subsection{Variational autoencoders}
\label{sec:VAE}
Variational autoencoders (VAE) are generative models that learn probabilistic mappings between a data space, where samples usually have a very complex distribution, and a latent space of well-distributed variables, \ie simple to sample from. 
Here, we will focus on the case of image data sets but VAEs have been successfully applied on various types of data \citep[for examples, see][]{2019arXiv190307137W,2018arXiv180604096R}.

More formally, we consider image samples denoted $x$ and latent variables denoted $z$ with a known prior $p(z)$ chosen to be a standard multivariate gaussian \citep*[a common choice in the literature, see][for examples]{2013arXiv1312.6114K,2017arXiv170202390S}. The generative model is a conditional distribution $p_\theta(x|z)$ (the likelihood) parametrised here by a convolutional neural network (CNN) with weights $\theta$ called the \textit{decoder}, that computes parameters of this distributions from input latent variables $z$. The VAE then introduces a second neural network, the \textit{encoder}, with weights denoted $\phi$, to approximate the intractable posterior $p_\theta(z|x)$ by another distribution $q_\phi(z|x)$.

The VAE then maximises the evidence of the training sample $p_\theta(x)$, which cannot be computed exactly, but can be bounded from below,
\begin{align}
    \centering
     \log p(x) \geq - D_{\rm KL} (q_\phi(z|x)||p(z)) + \mathbb{E}_{q_\phi}(\log p_\theta(x|z)), 
    \label{eq:loss_function_vae}
\end{align}
which defines the variational loss function of the VAE.
This loss receives two contributions. The first one is the Kullback-Leibler (KL) divergence between the approximate posterior and the prior, which ensures the gaussian distribution of the latent variables and acts as a regularisation term. The second one is given by the expectation value of the log-likelihood and is a reconstruction term. This term requires sampling latent variables according to the approximate posterior specified by the encoder. The denoising VAE only differs by the fact that the input $\tilde{x}$ is a noisy version of the target image $x$. In that case, $\tilde{x}$ is used in the regularisation term (as the encoder input) and $x$ is used in the reconstruction, but the variational loss does not change otherwise \citep{2015arXiv151106406J}.

\subsection{Implementation and architecture}
\label{sec:architecture_VAE}
In our implementation, the approximate posterior $q_\phi(z|x)$ is modelled by a product of univariate Gaussians with mean $\mu_\phi(x)$ and variance $\sigma^2_\phi(x)$, computed by the encoder (with trainable weights $\phi$), so that the KL divergence can be computed analytically. The decoder then maps latent variables $z$ onto the expectation values of independent Bernoulli distributions (in each pixel and for each band) which is interpreted as the output image to be compared to the target. We acknowledge that this is technically an abuse as the Bernoulli distribution is formally defined for binary variables rather than continuous variables in the range [0,1] where our rescaled images live (see \cref{sec:batch_generation}) and it formally breaks the variational objective in \cref{eq:loss_function_vae}. However, this is an empirical choice that is commonly found in the literature, including highly cited papers \citep{2013arXiv1312.6114K,2015arXiv151209300B, 2016arXiv160202282K,2016arXiv161105148J, 2016arXiv161102648D} and online tutorials, and ultimately justified by the performance of our method. We refer the reader to \citet{2019arXiv190706845L} for a detailed study of the consequences of this implementation and \cref{sec:discussion} for a discussion about possible improvements. 

\begin{figure}
	\includegraphics[trim={0.5cm 1cm 0.5cm 0cm}, width=\columnwidth]{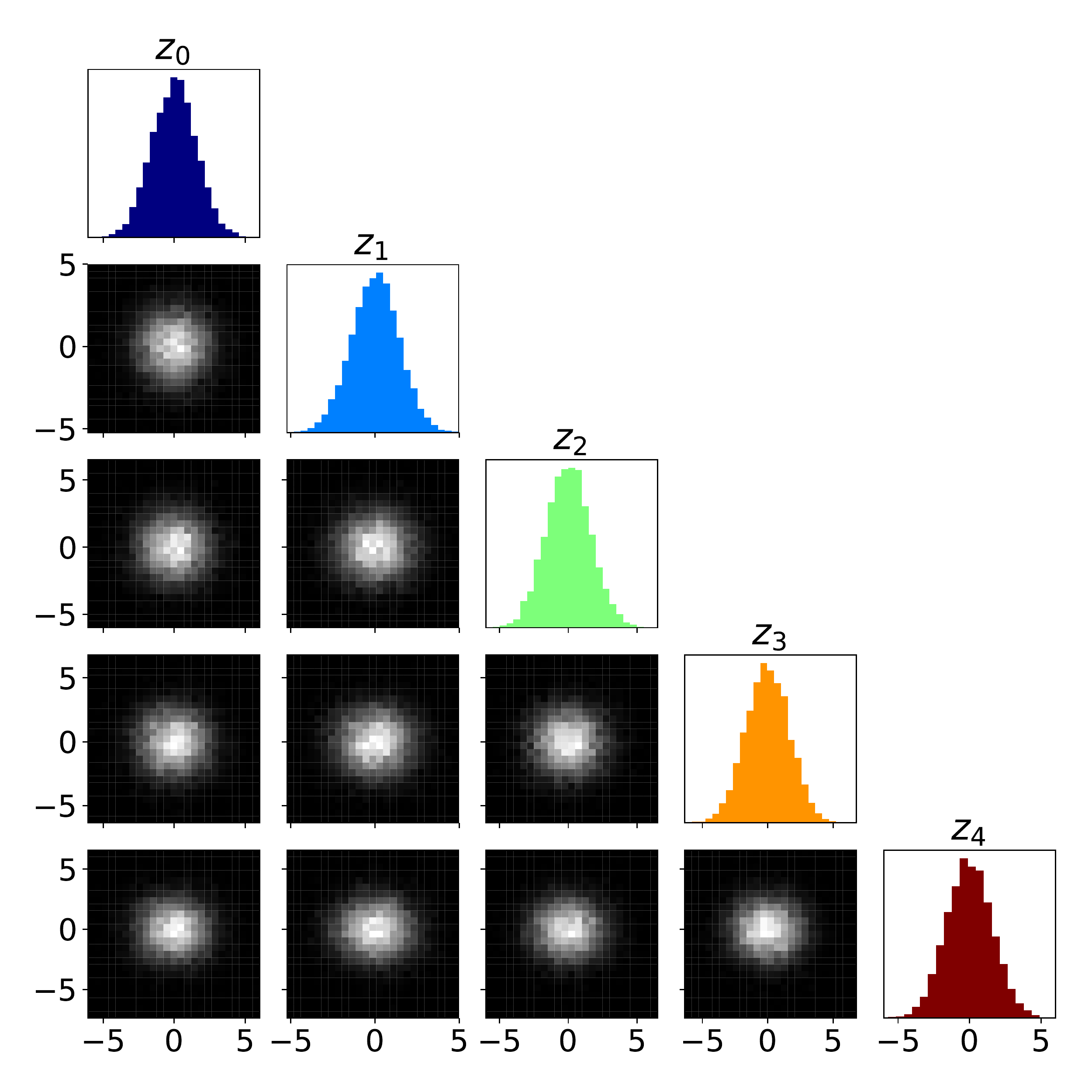}
    \caption{Corner plot of the five first dimensions of the latent space of a VAE trained on images composed of the LSST $ugrizy$ filters. All components show similar trends.}
    \label{fig:corner_latent}
\end{figure}

In order to optimise the reconstruction performance, we used a $\beta$-VAE \citep{Higgins2017betaVAELB}, \ie we minimised the weight of the Kullback-Leiber divergence in the loss by scaling it with a coefficient of $\beta$. Without this modification, the regularisation term prevails over the reconstruction term and prevents the network to learn to reproduce the input, a longstanding issue with vanilla VAEs.
In principle, this further entails the loss of the generative property of our model as this modification relaxes the regularisation on latent variables. However, we found that a value $\beta=10^{-2}$ yields acceptable reconstruction quality while, given the flexibility of the model, allowing latent variables to have a distribution very close to the Gaussian prior as shown on \cref{fig:corner_latent} (that could be further improved with regaussianisation techniques).

Our implementation is based on the Keras library\footnote{\url{https://keras.io/}} and the TensorFlow framework\footnote{\url{https://www.tensorflow.org/}} and uses a sequential network architecture presented in \cref{fig:VAE_deblender_architecture} with four sets of two convolutional layers in the encoder and symmetrically in the decoder (with transposed convolutional layer). In each set, both layers have the same number of filters but they have strides of, respectively (1,1) and (2,2), which allows us not to use pooling layers and thus reduces information loss, while downgrading the image. The convolutional layers have respectively 32, 64, 128 and 256 filters and all use 3x3 kernels. We added hidden dense layers of 256 units in both the encoder and the decoder. All these layers use parametric rectified linear unit (PReLU) activations. Finally, the encoder ends with two parallel dense layers ($\mu_\phi(x)$ and $\sigma^2_\phi(x)$) with 32 units, corresponding to the dimension of the latent space. In total, the encoder and the decoder have respectively \num{2.5} million and \num{3.3} million trainable parameters for the models trained on LSST $ugrizy$ bands and on all LSST and Euclid bands.

\subsection{Deblender}
\label{deblender}

The architecture of the deblender network is chosen to be the same as that of the VAE for simplicity. Before training, we will load decoder weights $\theta$ from a pretrained VAE and hold them fixed during training, \ie only the encoder weights $\phi^\prime$ are trainable.
The encoder learns a mapping from blended images to the posterior parameters $\mu_{\phi^\prime}$ and $\sigma_{\phi^\prime}$ used to draw latent variables. Those are then used as input to the fixed decoder, which outputs the likelihood parameters, needed to evaluate the reconstruction term in the loss, together with the target noiseless image of the isolated central galaxy.
Therefore the encoder learns to separate the central galaxy from every other part of the input image (noise and neighbouring galaxies) and to map it into the latent space. The encoding must be done in a way that permits the decoder to reproduce the centred galaxy accurately. As a consequence, our method can be considered as operating deblending in latent space with a mapping created from a prior generative model of isolated galaxies.

For the deblender, we also keep the Kullback-Leiber divergence and the reconstruction term in the loss, which reads
\begin{equation}
\begin{split}
    \centering
    L(\phi^\prime, x_{\rm in}, x_{\rm target}) = & \beta\,D_{\rm KL} \qty(q_{\phi^\prime}(z|x_{\rm in})||p(z)) \\
    &- \mathbb{E}_{q_{\phi^\prime}}\qty[\log p_\theta(x_{\rm target}|z)],
\end{split}
    \label{eq:loss_function_vae_deblender}
\end{equation}
but here we minimise over the mapper weights $\phi^\prime$. Here $x_{\rm in}$ is the noisy input and blended image and $x_{\rm target}$ is the noiseless target image of the central galaxy. Note that we tried to initialise those weights either randomly or from the trained encoder and obtained the same results.

\section{Results}
\label{sec:results}

\subsection{Reconstruction metrics}
\label{metrics}
Weak lensing is our main focus in this work and we thus assess the performance of our method by measuring how well shape parameters and magnitudes can be recovered by the deblender (see definitions below).
To do so, we first measure ellipticities and magnitudes of the noiseless, denormalised, target images in the test sample, described in \cref{sec:data_sample}. Then, when training is finished, we run the deblender on the test sample and repeat those measurements on the denormalised output of the deblender. Finally, we compute errors on ellipticities and magnitudes between the target and output images and use them to compute performance metrics.
We present the distribution of the measured errors as functions of other relevant quantities, such as signal-to-noise ratio and blendedness metrics, in order to gain intuition about the assets and limitations of our method. More precisely, we measure:
\begin{enumerate}
    \item the PSF-corrected ellipticity defined as the reduced shear estimator $|e| \equiv (a-b)/(a+b)$, where $a$ and $b$ are the semi-major and semi-minor radii. The measurement is performed in the LSST $r$-band from PSF-convolved images using the \verb|HSM| module in \galsim and the \verb|EstimateShear()| function with the Kaiser-Squires-Broadhurst method \citep*[KSB,][]{1995ApJ...449..460K}, to which we provide the fixed PSF used to generate the images (see \cref{sec:data_sample}).
    \item the magnitude in the $r$-band, computed from the total flux, itself obtained by simply summing the number of photons from every pixel. We have verified that we obtain very similar results in all bands.
\end{enumerate} 
In order to demonstrate and quantify the benefits of using multiple bands and multiple instruments, we repeat the training of both networks and the analysis for images consisting of
\begin{enumerate}
  \item the six LSST bandpass filters ($ugrizy$, 6 bands), and
  \item all LSST and Euclid bandpass filters together (10 bands).
\end{enumerate}

Finally, we also evaluate the robustness of our networks to signal-to-noise ratio (SNR) as it is expected to have a significant impact on the performance. We define the SNR as
\begin{equation}
    S/N = \sqrt{\sum_{p \in \rm pixels}\frac{I_p^{2}}{\sigma^2(I_p)}},
\end{equation}
where $I_p$ is the intensity of the signal in pixel $p$ in the $r$-band and the variance $\sigma^2(I_p)$ is the sum of the signal and the sky background. This ratio is measured on isolated galaxy images and this definition is therefore valid for individual objects. Nevertheless, we use the same values when objects are blended for our shape and magnitude reconstruction tests, which over-evaluates the signal-to-noise ratios that could be reached. Therefore, figures concerning deblending performance involving SNR splits present conservative results.  

\subsection{Prior for single galaxies with a VAE}
\label{sec:vae_results}

The first network, the \textit{prior model}, is intended to learn a generative model of multi-band images of isolated galaxies. As a probabilistic model, the output is conditioned on a random draw in latent space, therefore we do not compare measurements on individual galaxies but rather the statistical properties of errors on the test sample.

\begin{figure*}
    \centering
    \includegraphics[trim=0 2.6cm 0 3cm,clip, width=17cm]{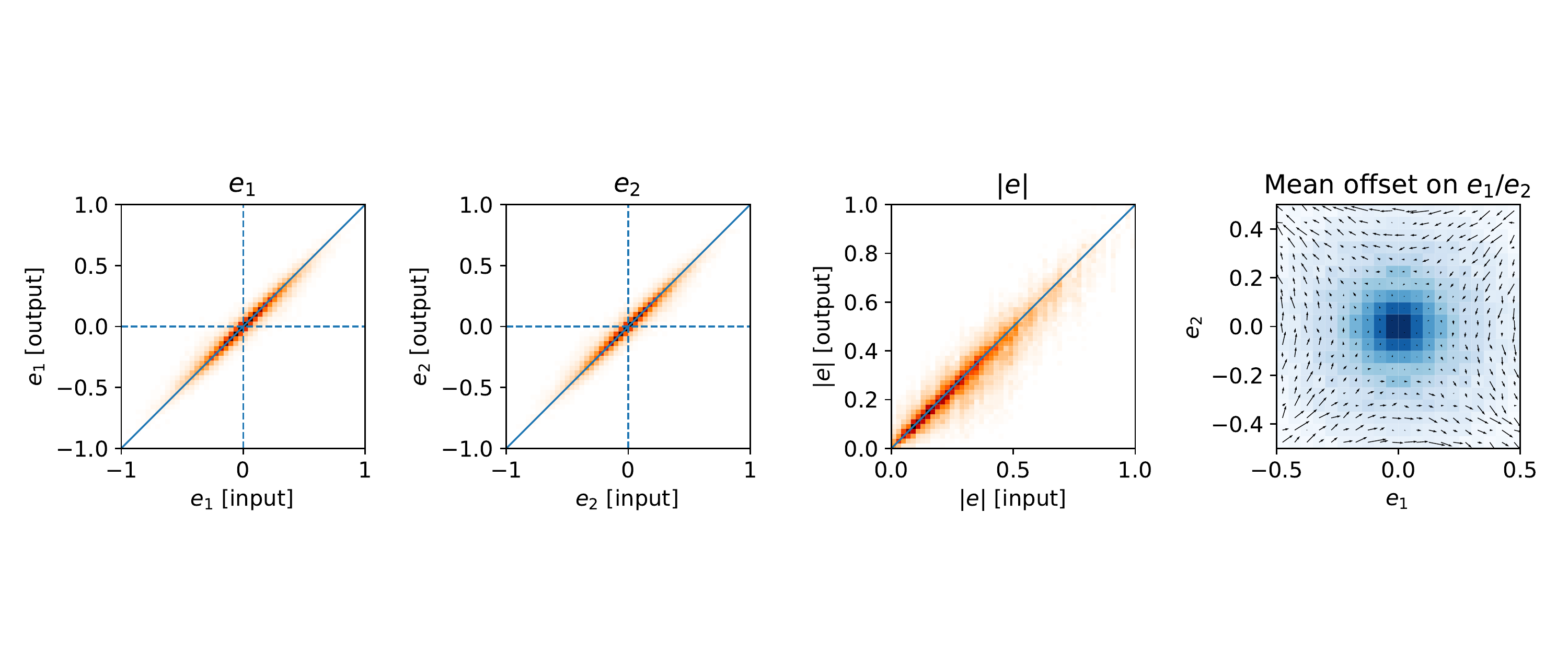}
    \caption{Distributions of output shape parameters as a function of parameters measured on target images ($e_1$, $e_2$ and $|e|$ from the leftmost panel to the third one). The rightmost panel shows the mean offset on $e_1$ and $e_2$ in the $(e_1,e_2)$ plane with arrows pointing to the average output values of these components (with the same scale as the axes).}
    \label{fig:vae_e_repro}
\end{figure*}

\begin{figure*}
    \includegraphics[width=\columnwidth]{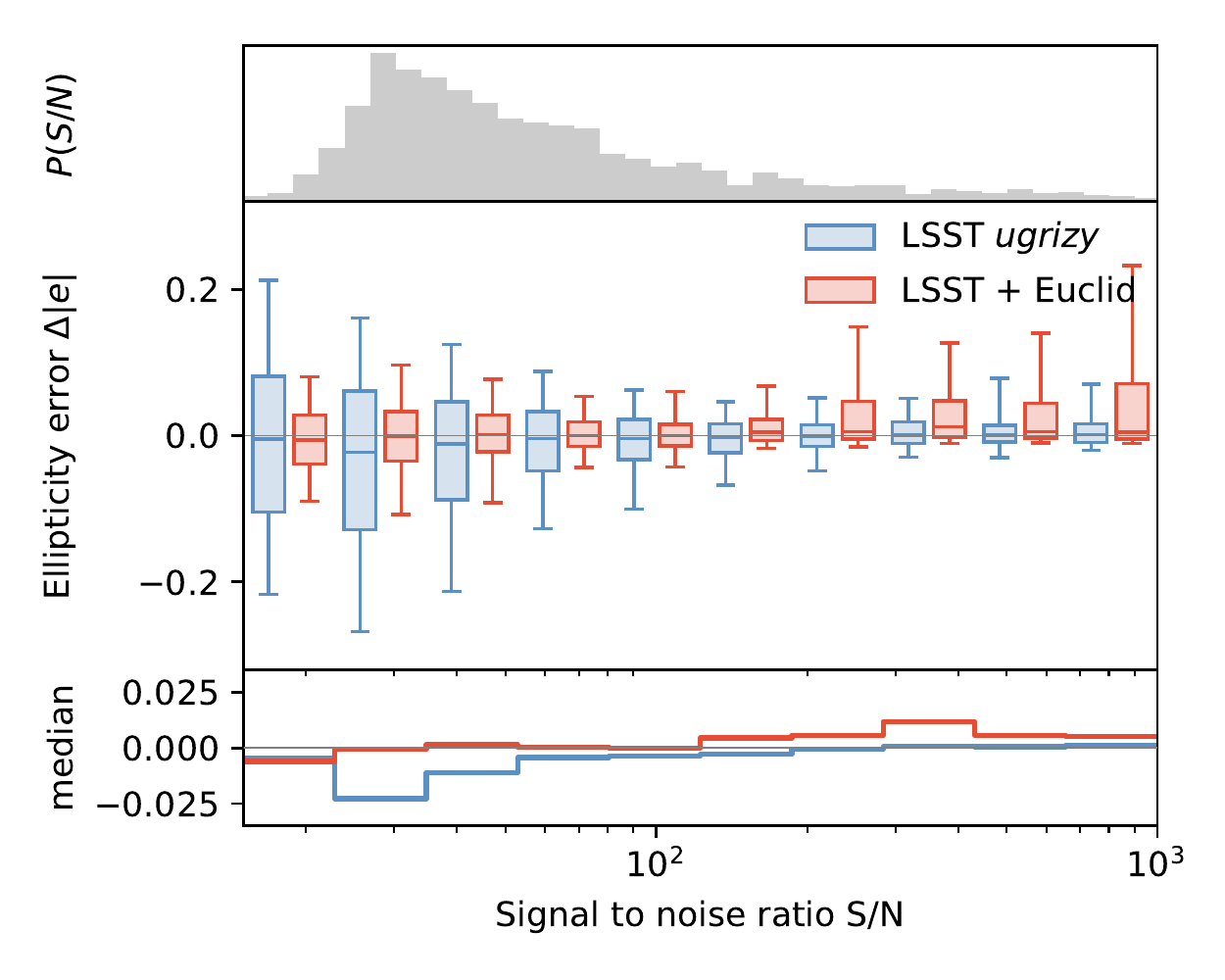}
    \includegraphics[width=\columnwidth]{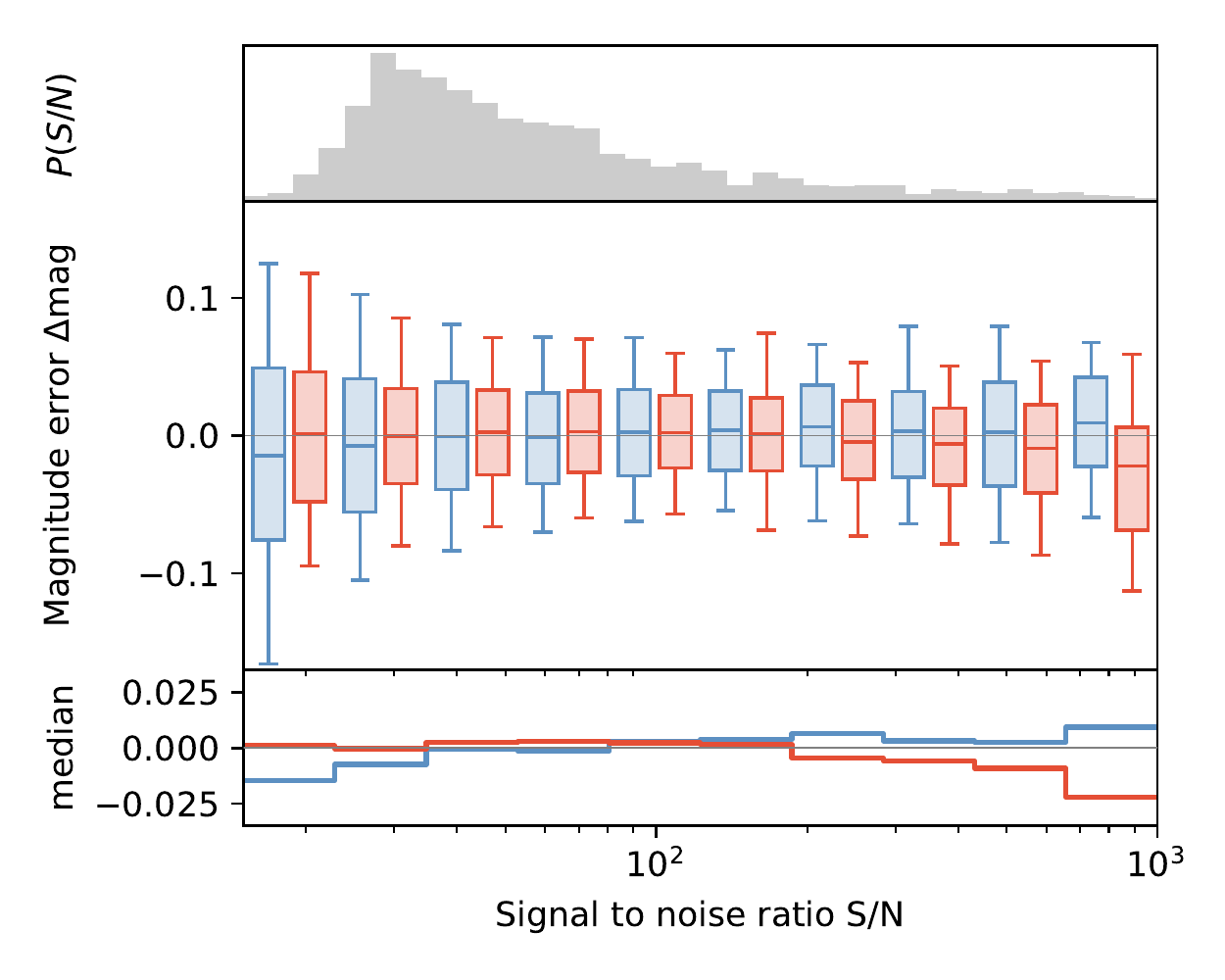}
    \caption{
    Distributions of differences in absolute observed ellipticity $|e|$ (left) and $r$-band magnitude (right) between target and output images from the test sample. Results for the VAEs trained on LSST bands only are shown in blue and those obtained when including Euclid bands are shown in red. The distributions are computed on subsets of the test sample split in bins of signal-to-noise ratio $S/N$ (left) and $r$-band magnitude (right).
    In each plot, the top panel shows the distribution over the test sample of the quantity on the x-axis, the middle panel shows the distribution of errors when the test sample is split in bins of this particular quantity, and the bottom panel shows the median error. In the middle panel, the boxes show the median and $\pm1\sigma$ percentiles and the whiskers show the $\pm2\sigma$ percentiles.
    }
    \label{fig:results_vae}
\end{figure*}

In \cref{fig:vae_e_repro}, we first show errors for both ellipticity components $(e_1, e_2)$ in the prior model, finding no particular difference in reconstruction between the ellipticities along the pixel grid ($e_1$) and at \SI{45}{\degree} ($e_2$), which is not guaranteed \textit{a priori}. This figure also shows on the right the mean 
offset on $e_1$ and $e_2$ in the plane $(e_1,e_2)$ with arrows pointing to the output values of these components (with the same scale as the axes). We find offsets to be very small around the centre (round galaxies) and to increase for more elliptical galaxies, which are also less represented in the training sample.
More surprisingly, we find an asymmetry in the $(e_1, e_2)$ plane for more elliptical galaxies, and that the network outputs galaxy images with ellipticities slightly biased towards the directions $(-1,1)$ and $(1,-1)$. Note, however, that results are symmetric by swapping components.

Since we have shown that the network has similar reconstruction error on the two components, we now focus on the errors on the absolute ellipticity $|e|=\sqrt{e_1^2+e_2^2}$ and $r$-band magnitude, in particular on the median error and the width of the distributions. \Cref{fig:results_vae} shows those distributions in both configurations when splitting the test sample in ten bins of SNR. Boxes show the median and percentiles corresponding to $\pm1\sigma$ of a gaussian, the whiskers show $\pm2\sigma$ percentiles.
We find the distributions of ellipticity errors to be centred around zero with small deviations as the median is contained within $\pm\num{0.025}$. Increased spread around the median arises at SNR typically below 50, particularly for the VAE trained on LSST bandpass filters only, which is likely due to the blurring of object edges in the presence of noise. The network produces on average an output image that is slightly rounder (closer to the average image) than the input image, which creates a small negative bias. The median and spread then reduce as SNR increases, even in the high SNR region which is sparsely populated.
We find similar trends for both tested configurations, although the median and spread of the error distributions decrease by, respectively, 22\% and 47\% on average, when using Euclid bandpass filters (multi-instruments approach) in addition to the six LSST bandpass filters. 
The improvement is even more significant at low SNR where the median (respectively spread) is reduced by 81\% (59\%) on average for galaxies with $S/N < 100$.
The right panel of \cref{fig:results_vae} presents the error distributions on the reconstruction of the $r$-band magnitude as a function of SNR. We find that our models reproduce magnitudes with errors below 0.2 at $2\sigma$ and median within $\pm\num{0.025}$, reducing in the middle range of magnitudes and increasing for very bright objects that are outliers in the training sample. Adding Euclid bandpasses reduces the median and spread on the magnitude error by respectively 72\% and 21\% on average for galaxies with $S/N < 100$. However, on the entire sample, the spread on the magnitude error is reduced by 12\% but we do not observe any improvement on the median error, probably due to very bright outliers.

From these results, we conclude that our model is able to learn features from noisy images and accurately reproduce parameters relevant for weak lensing studies. They also show that the VAE is able to integrate information from observations in multiple bands and instruments to learn tighter posterior and likelihood distributions, leading to more accurate results across all bands.
It is likely that extra information provided by Euclid filters mainly comes from the VIS instrument, which collects information on a very broad bandpass with higher resolution than LSST (\SI{0.1}{\arcsecond}/pixel for VIS \textit{vs.} \SI{0.2}{\arcsecond}/pixel for LSST), yielding significant reductions of errors on galaxy shapes, especially at low SNR. 
These results are consistent with the findings of \citet{2019arXiv190108586S}, \ie that a pixel-level joint analysis of LSST and Euclid data yields significant improvements of shape measurements.


\subsection{Deblending performance}

\begin{figure*}
    \begin{minipage}[b]{.49\linewidth}
        \centering
        \includegraphics[trim={0 0.2cm 0 0.4cm}, clip, width=0.8\linewidth]{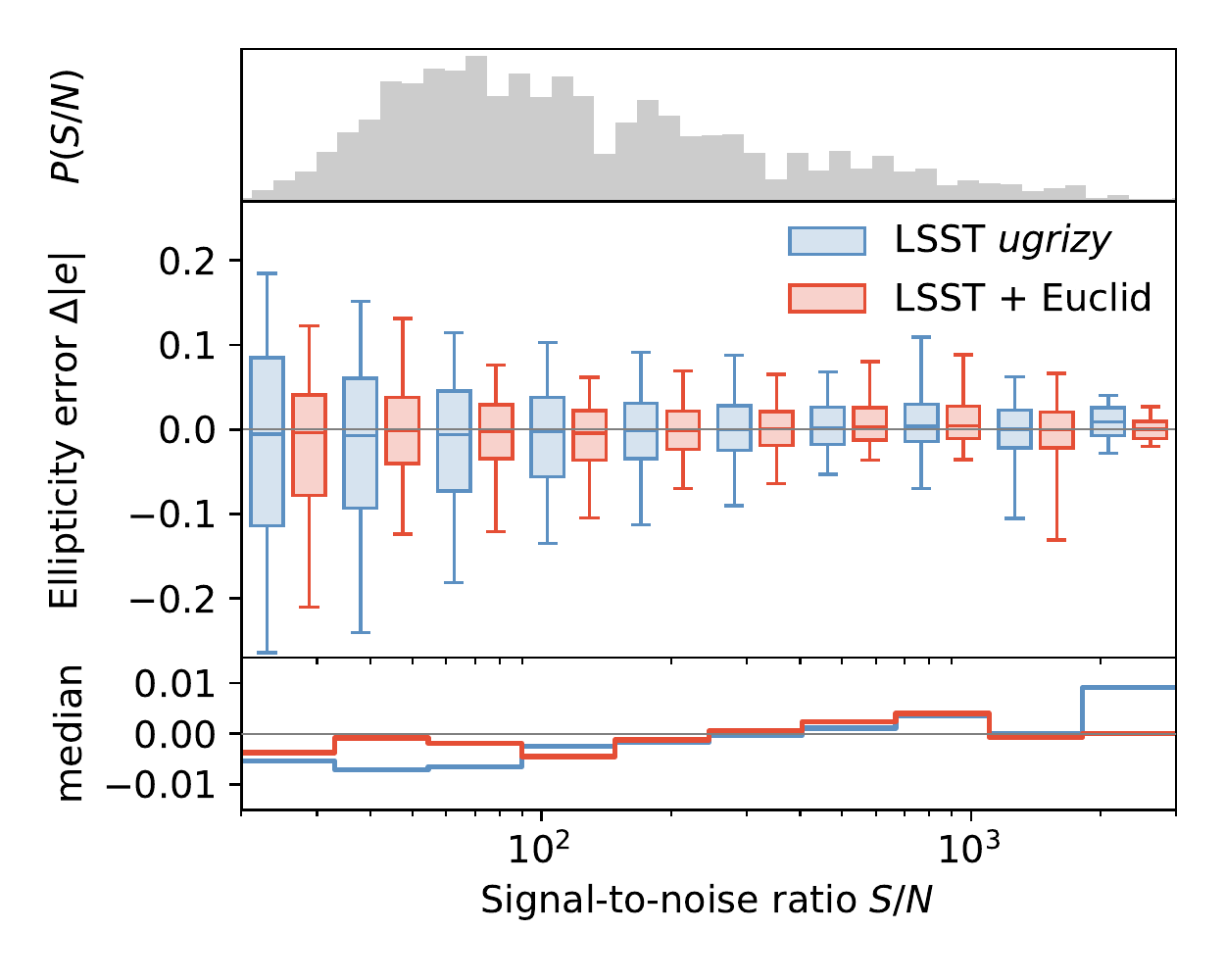}
        \includegraphics[trim={0 0.2cm 0 0.4cm}, clip, width=0.8\linewidth]{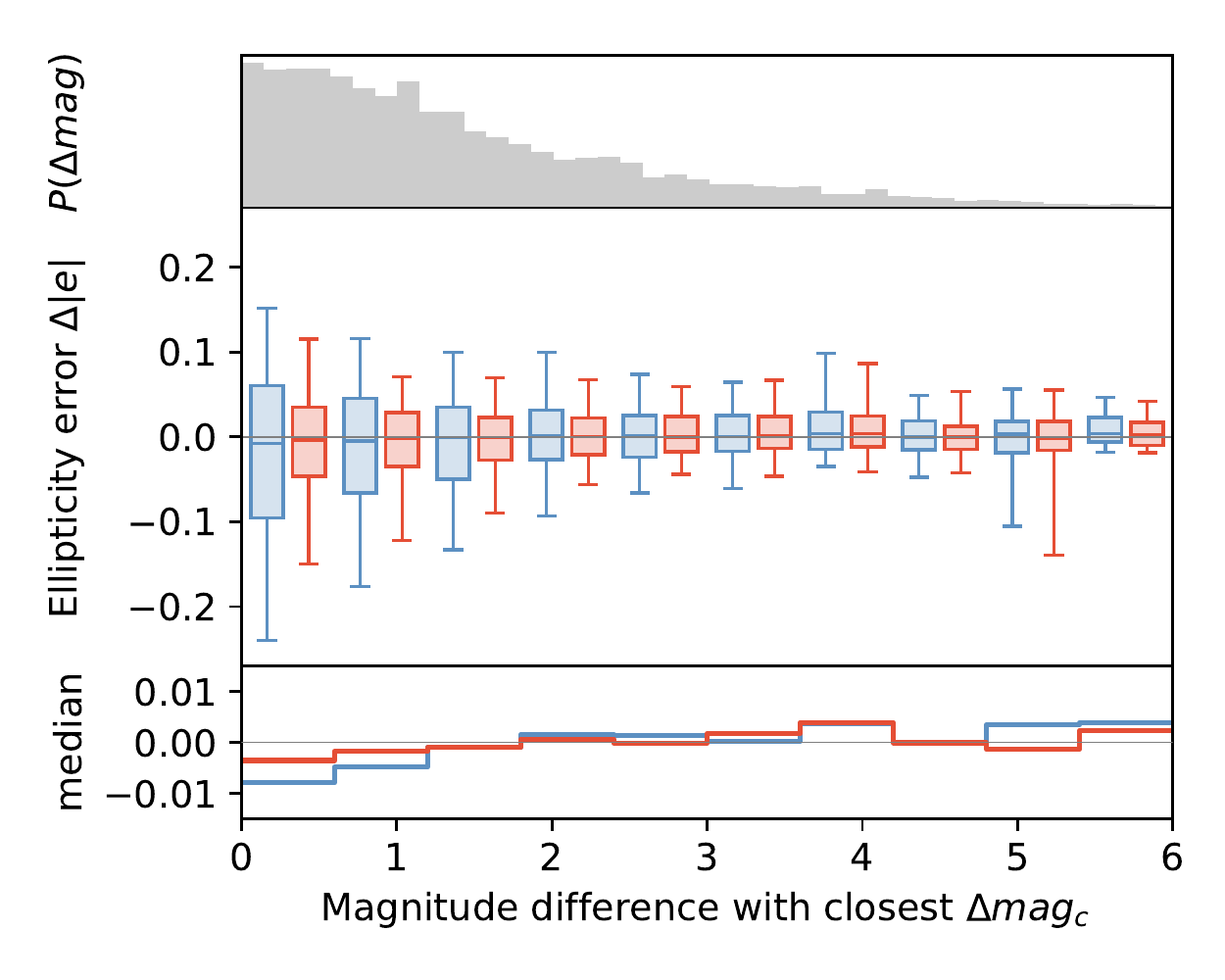}
        \includegraphics[trim={0 0.2cm 0 0.4cm}, clip, width=0.8\linewidth]{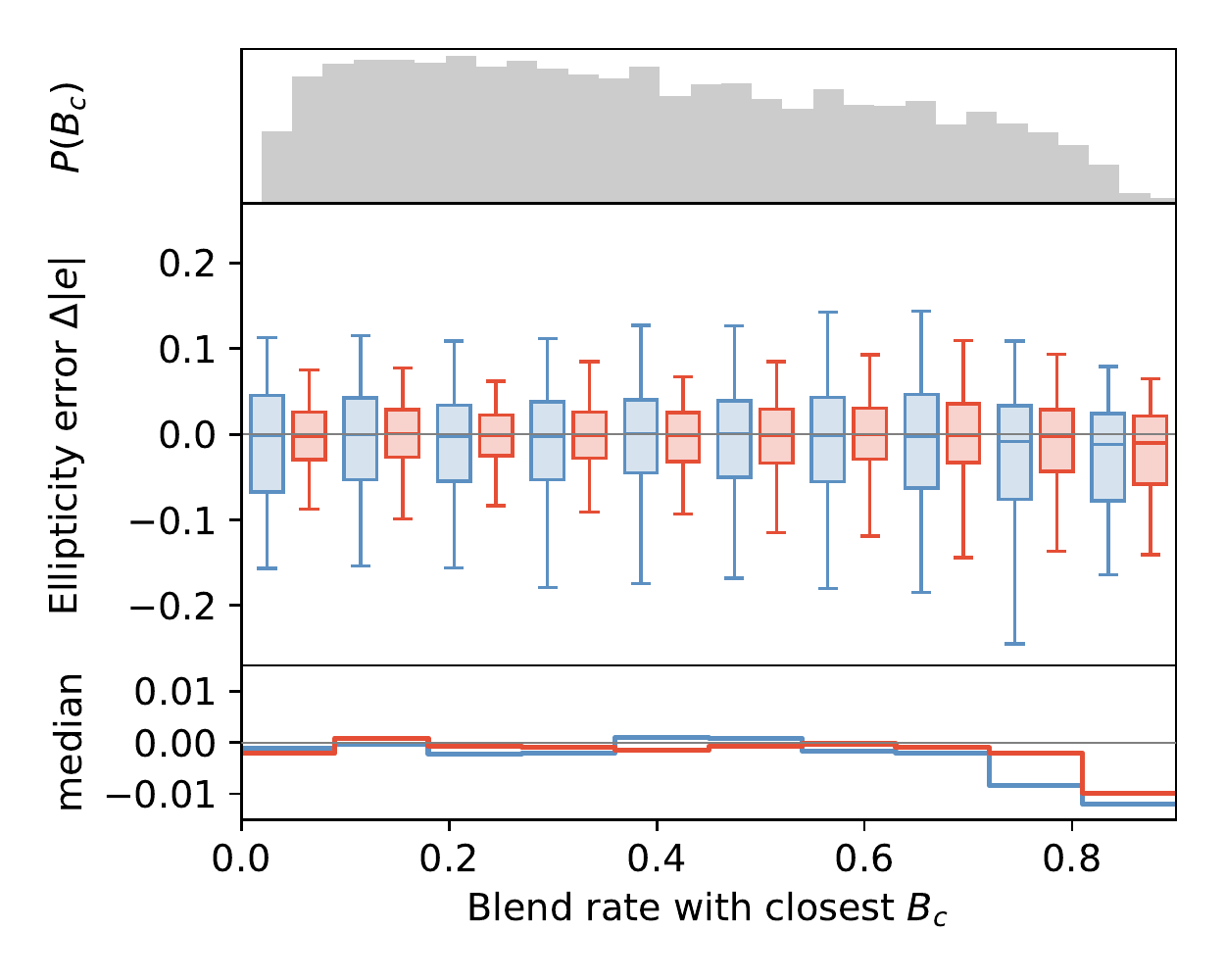}
        \includegraphics[trim={0 0.2cm 0 0.4cm}, clip, width=0.8\linewidth]{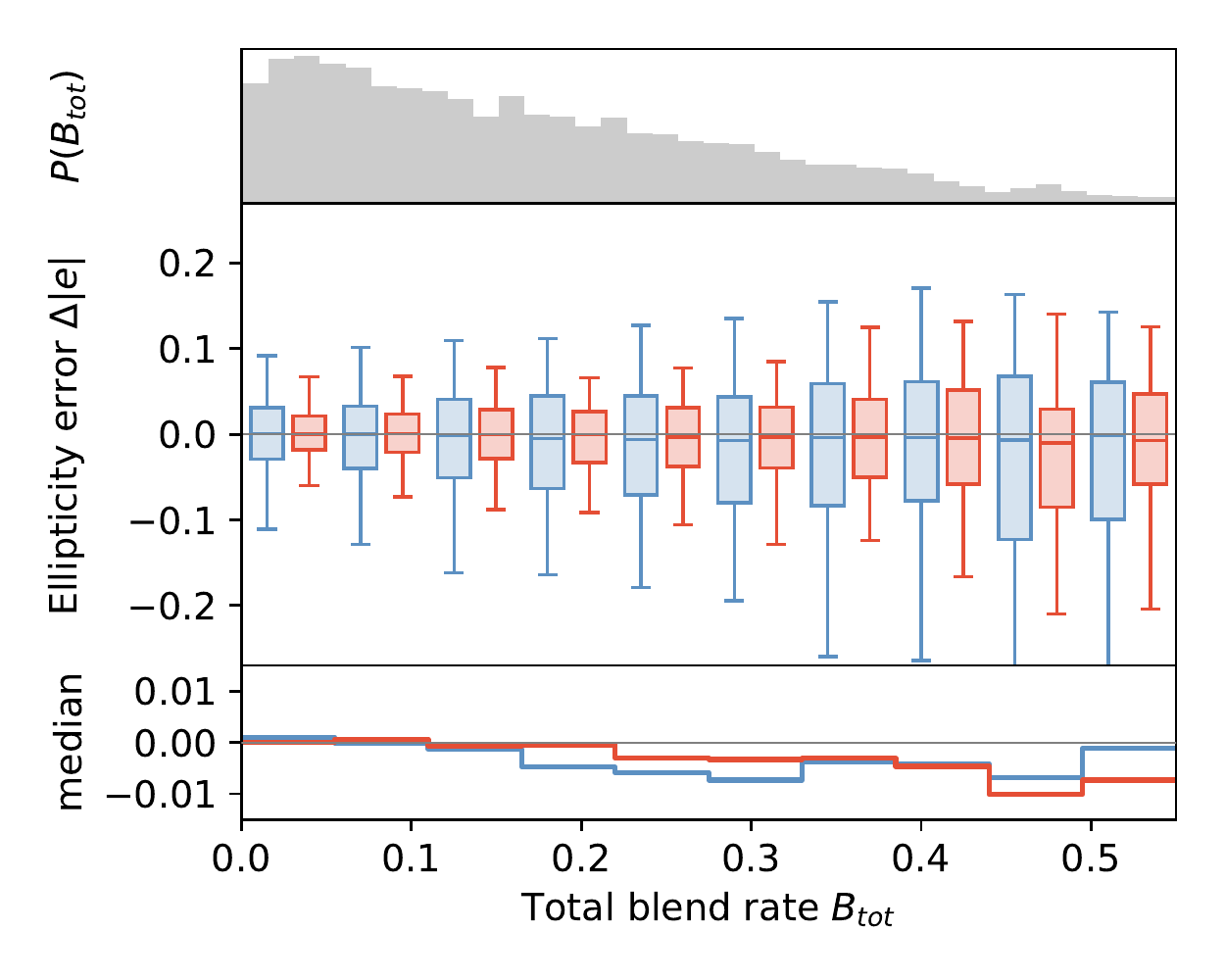}
        \text{a) Error in ellipticity}
    \end{minipage}
    \hfill%
    \begin{minipage}[b]{.49\linewidth}
        \centering
        \includegraphics[trim={0 0.2cm 0 0.4cm}, clip, width=0.8\linewidth]{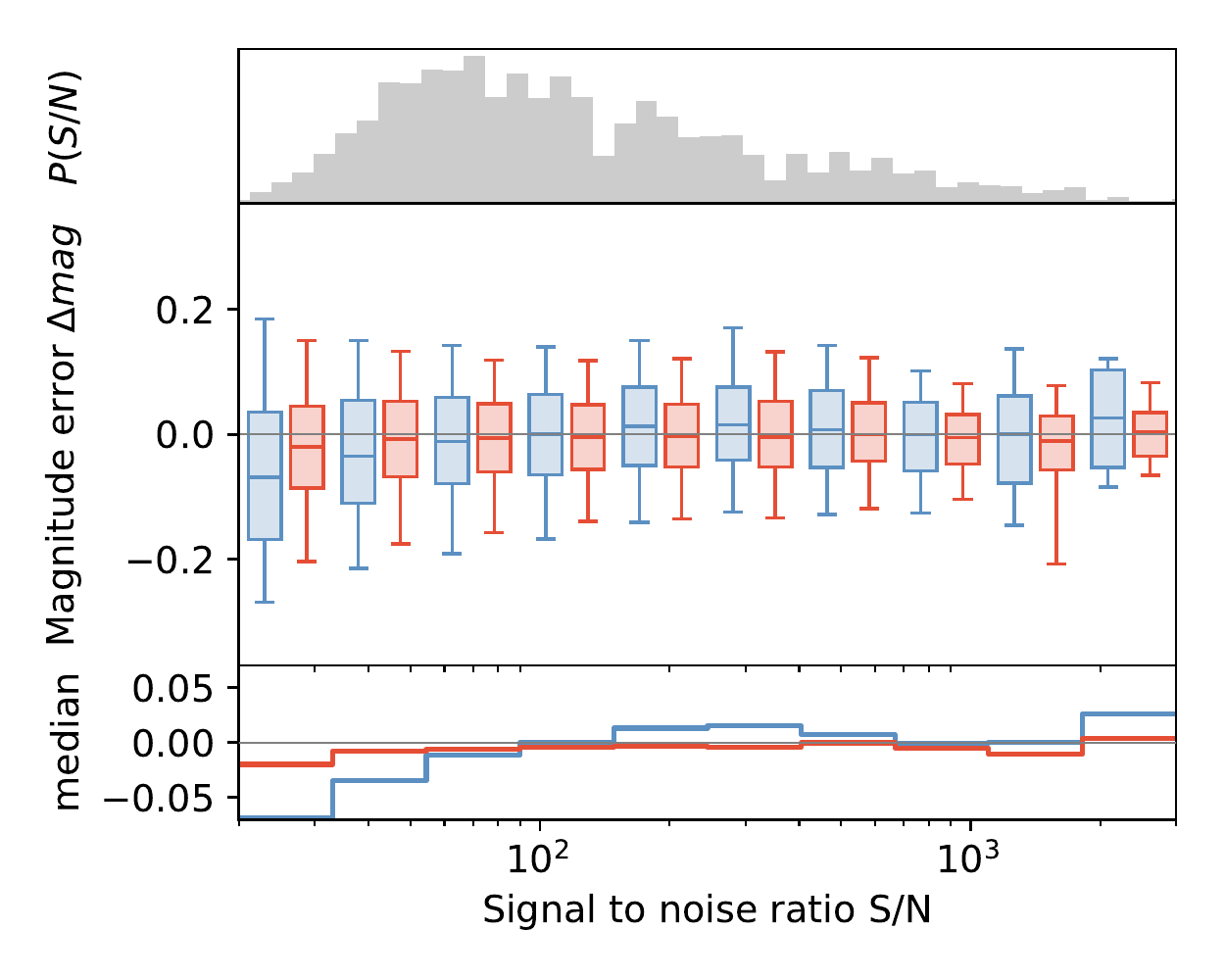}
        \includegraphics[trim={0 0.2cm 0 0.4cm}, clip, width=0.8\linewidth]{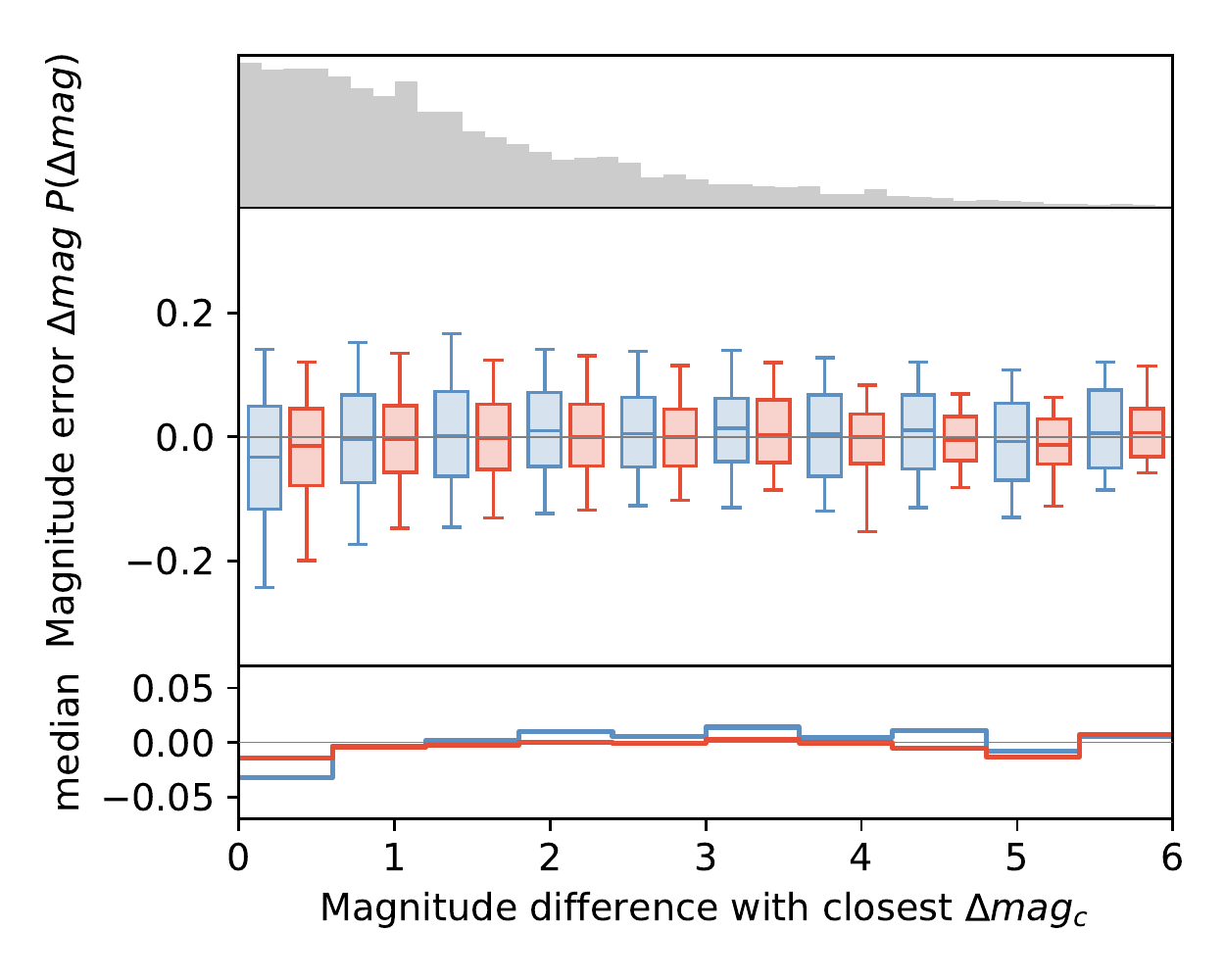}
        \includegraphics[trim={0 0.2cm 0 0.4cm}, clip, width=0.8\linewidth]{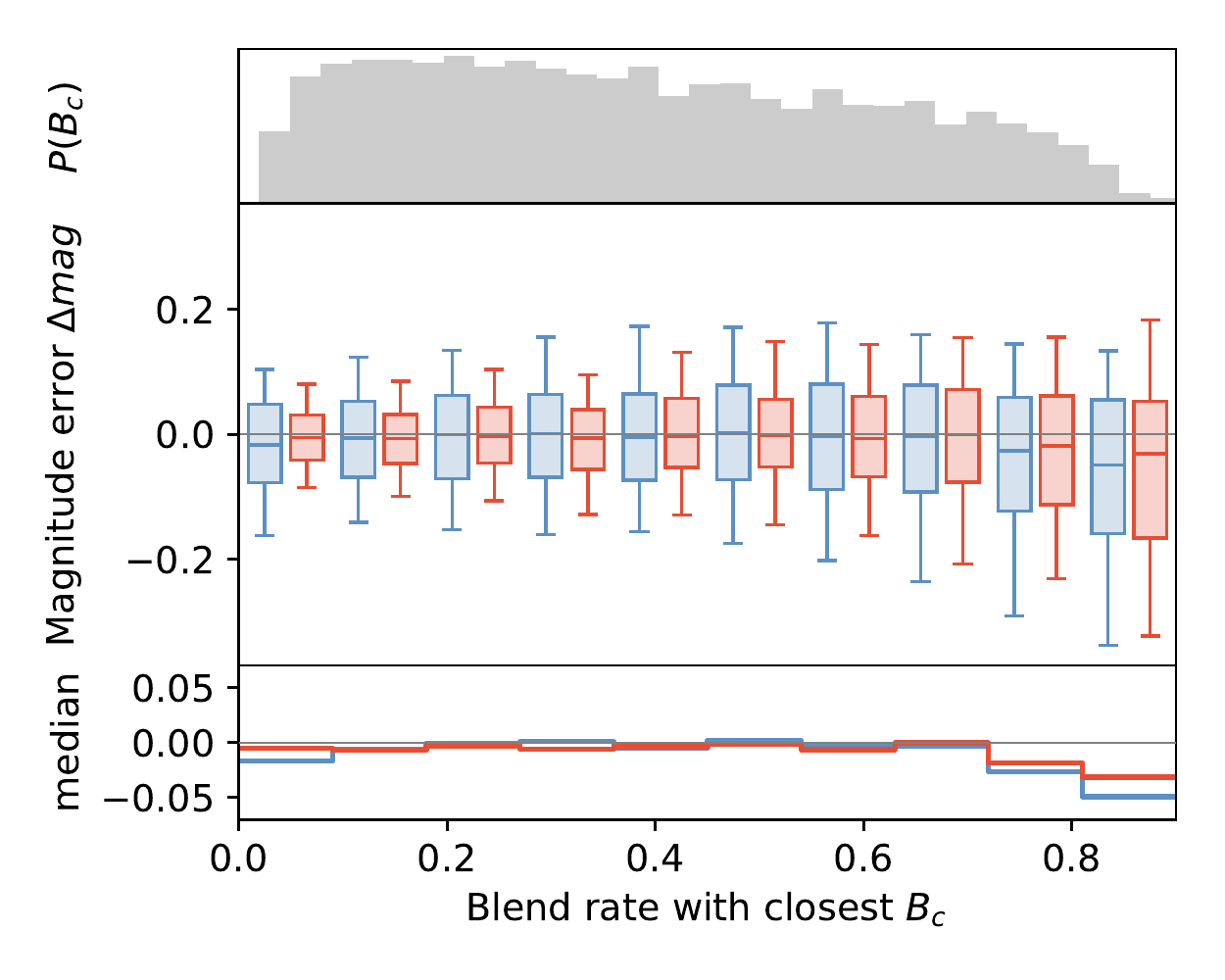}
        \includegraphics[trim={0 0.2cm 0 0.4cm}, clip, width=0.8\linewidth]{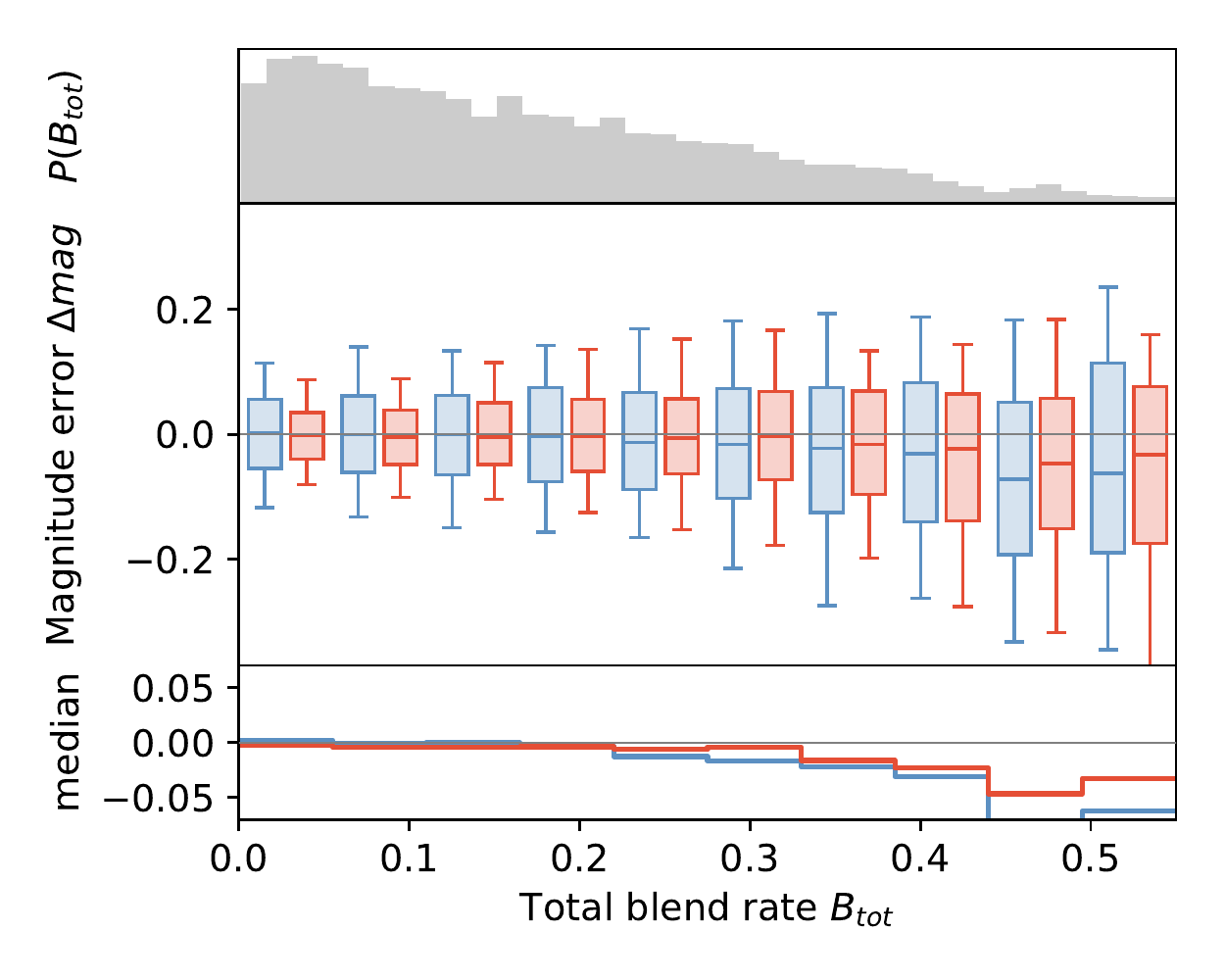}
        \text{b) Error in magnitude}
    \end{minipage}
    \caption{Distributions of errors on absolute ellipticity $|e|$ (left) and $r$-band magnitude (right) for the deblenders trained on LSST (blue) and LSST+Euclid (red) images. These distributions are shown as a function of the signal-to-noise ratio $S/N$ of the target galaxy, the magnitude difference with the closest galaxy $\Delta {\rm mag}_{\rm c}$, the blendedness with the closest neighbour $B_{\rm c}$ and the total blendedness $B_{\rm tot}$. In each plot, the top panel shows the distribution over the test sample of the quantity on the x-axis, the middle panel shows the distribution of errors when the test sample is split in bins of this particular quantity, and the bottom panel shows the median error. In the middle panel, the boxes show the median and $\pm1\sigma$ percentiles. The whiskers show the $\pm2\sigma$ percentiles.
    In each column, all plots have the same vertical scales to facilitate comparison.}
    \label{fig:deblender_output}
\end{figure*}

As explained in \cref{deblender}, we now instantiate a second neural network to perform deblending, with the same architecture, and reusing the generative model trained and tested in the previous section. The decoder weights are loaded from the trained prior models and held fixed during training so that only the deblender's encoder parameters are optimised.
In this section, we present results obtained on images using the same two previous configurations where target galaxies are perfectly centred. A random sample of the deblender's input and output images is shown in \cref{app:Deblending_images}.

\Cref{fig:deblender_output} shows the distributions of errors on target galaxy ellipticity and magnitude as a function of signal-to-noise ratio ($S/N$), magnitude difference with the closest neighbour galaxy ($\Delta{\rm mag_c}$) and the two blendedness metrics defined in \cref{sec:blendedness_metrics} ($B_c$ and $B_{\rm tot}$).
For both configurations, we find the median errors to be within $\pm\num{0.015}$ and $\pm\num{0.08}$ for ellipticity and magnitude respectively, across the entire test sample.
As expected, we find the largest median errors and spread at low SNR, small difference in magnitude with the closest neighbour and high blend rates. In these regimes, median errors are negative, meaning that the network produces rounder and brighter images, picking up flux from neighbouring galaxies.
However, median errors remain limited even in critical cases. For example, at blend rates $B_c$ of 80\%, median errors on ellipticity and magnitude are respectively below $0.008$ and $0.035$.

Comparing both configurations, we observe some significant improvements when including Euclid filters. The largest observed effect is the reduction of the median error on ellipticities by 40 to 47\% and of the spread by about 33\% on average, which is visible when the test sample is split in bins of SNR, magnitude difference and blend rate with closest neighbour (three top rows in \cref{fig:deblender_output}). A smaller improvement is observed when the data is split in bins of total blend rate $B_{\rm tot}$, with gains around 8\% on the mean and 36\% on the spread of the error distribution. Magnitude errors are also reduced across blend rates, SNR and magnitude difference: 27 to 63\% for the median and 17 to 33\% for the spread of the distribution.

We also characterise the impact of our deblending method on shear measurements. To do so, we generate five test samples of \num{10000} input and target images. Each sample is made of the same sets of galaxies (identical profiles, positions and rotations) and noise realisations, but with different amounts of constant shear applied, namely $(0,0)$, $(+0.01,0)$, $(-0.01,0)$, $(0,+0.01)$ and $(0,-0.01)$. We decompose the ellipticities measured on output images $e_i^{\rm out}$, for each component $i=1,2$, as a function of the target image ellipticity (considered as the intrinsic ellipticity of the galaxy) and of the true shear $\gamma$ applied to the sample, as
\begin{equation}
    e_i^{\rm out} \approx (1+\alpha_i) e_i^{\rm in} + (1+m_i) \gamma_i + c,
    \label{eq:ellipticity_shear_bias}
\end{equation}
where $\alpha_i$ is the mean ellipticity bias when zero shear is applied (its measurement is presented in \cref{sec:decentring_results}), $m_i$ is the multiplicative shear bias, and $c$ is the additive shear bias. In order to measure the shear biases $m_i$, we start by computing the average ellipticity, denoted $\expval{e_i^{\rm out}}_\pm$, of the samples where a constant shear $\gamma_i=\pm0.01$ has been applied (with the other component unsheared).
Shear biases can then be obtained by differentiating \cref{eq:ellipticity_shear_bias}, where intrinsic input ellipticities $e_i^{\rm in}$ are equal for all samples, leading to
\begin{equation}
    {m_i \approx \frac{\expval{e_i^{\rm out}}_+ - \expval{e_i^{\rm out}}_-}{\Delta\gamma}}-1
\end{equation}
with $\Delta\gamma=0.02$. This method resembles the metacalibration algorithm \citep{2017arXiv170202600H,2017ApJ...841...24S} to quantify shear responses, except that we apply shear directly to the image model and not to a noisy image. We find uncalibrated multiplicative shear biases of the order of 1 to 4\% for both LSST and LSST+Euclid configurations on deblended images. These values are to be compared with biases of 4 to 6\% measured on target images, characterising the bias due to the shear estimator alone on our test sample. We interpret this result as a percent-level negative bias introduced by the deblending method, consistent with previous observations. The observed additive bias, measured on the $(0,0)$-sheared sample, is consistent with zero (up to the sampling error). Note that these are uncalibrated biases, measured on a sample of highly blended galaxies scenes, unrepresentative of the full sample. As such, they are not directly comparable to the values set by the LSST Science Requirements Document \citep{2018arXiv180901669T}.
Nevertheless, following the principle used here for this measurement, the calibration of this multiplicative bias could be achieved with the metacalibration algorithm, which we do not address any further in this work.

\subsection{Effect of decentring}
\label{sec:decentring_results}

\begin{figure*}
    \begin{minipage}[b]{.49\linewidth}
        \centering
        \includegraphics[trim={0 0 0 0.5cm},  width=0.88\linewidth]{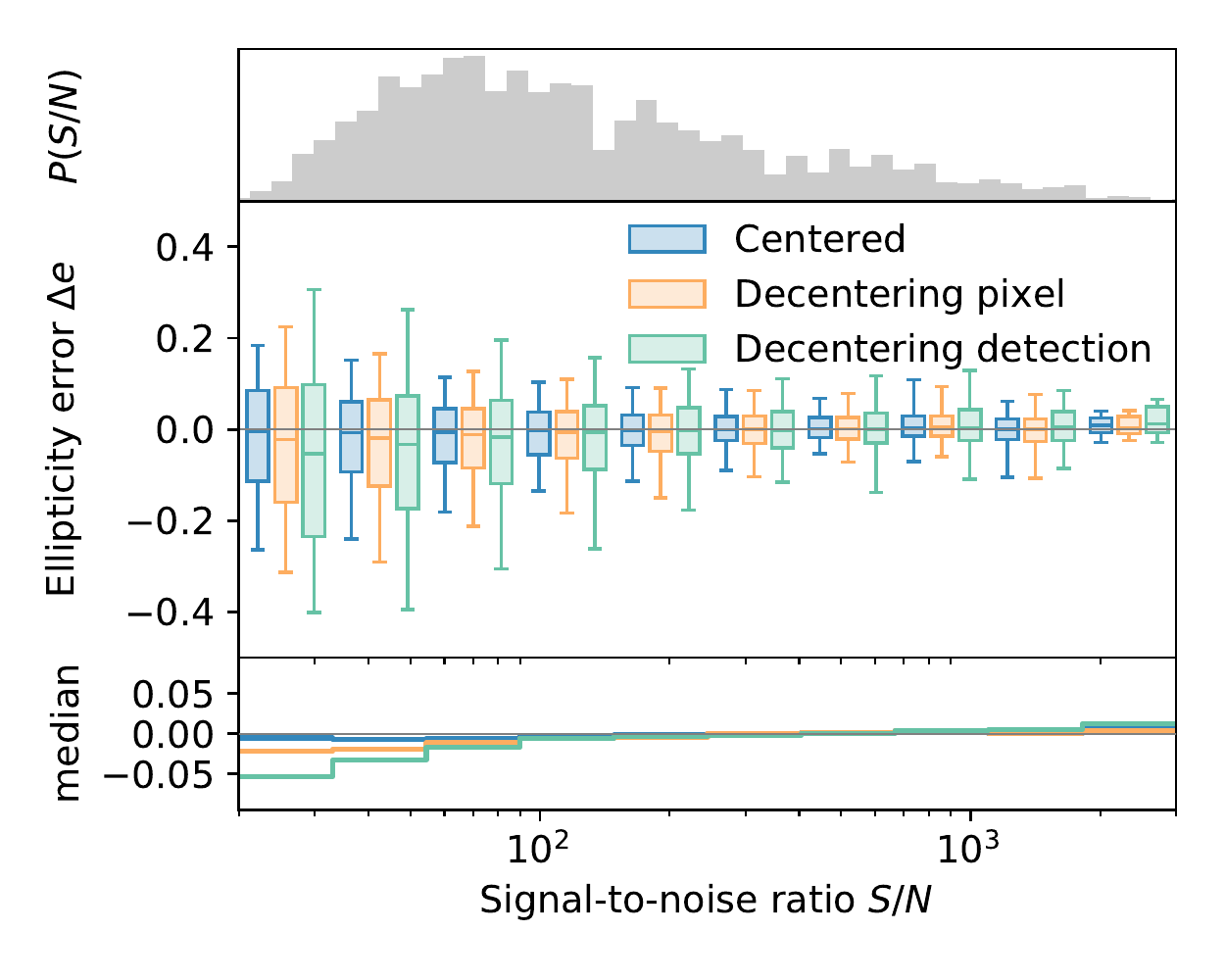}
        \\
        \includegraphics[trim={0 0 0 0.5cm}, width=0.88\linewidth]{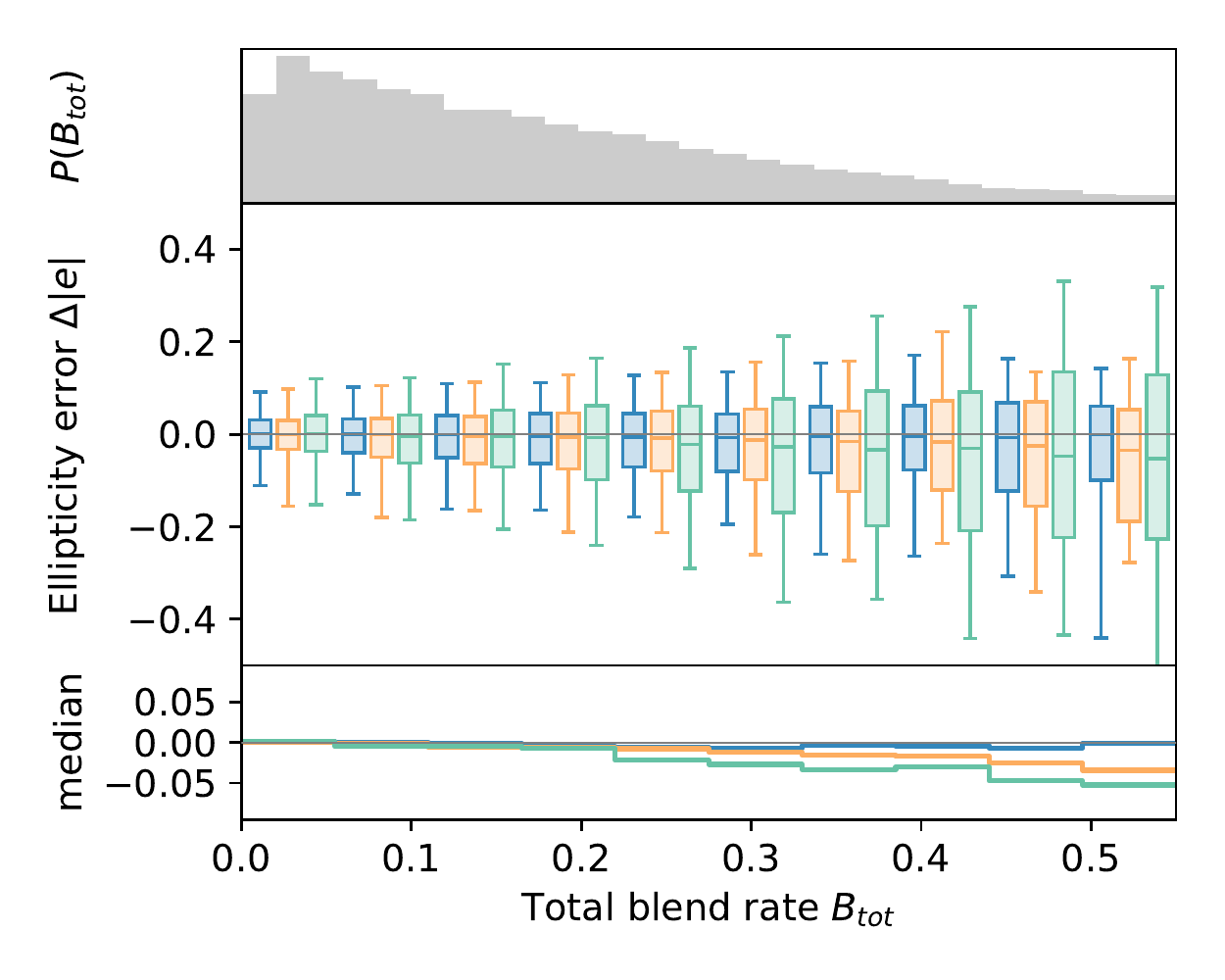}
        \\
        \text{a) Error in ellipticity}
    \end{minipage}
    \hfill%
    \begin{minipage}[b]{.49\linewidth}
        \centering
        \includegraphics[trim={0 0 0 0.5cm}, width=0.88\linewidth]{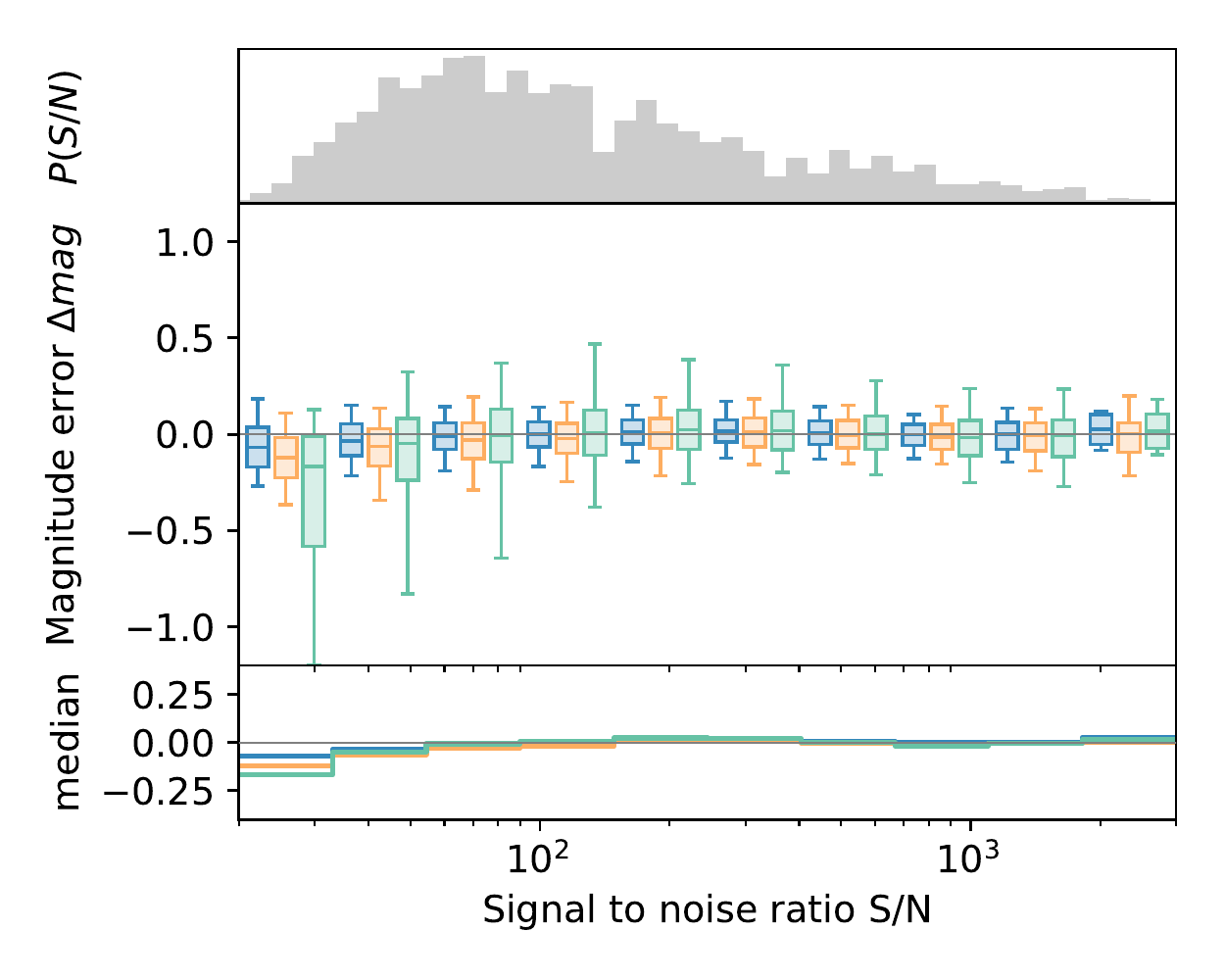}
        \\
        \includegraphics[trim={0 0 0 0.5cm}, width=0.88\linewidth]{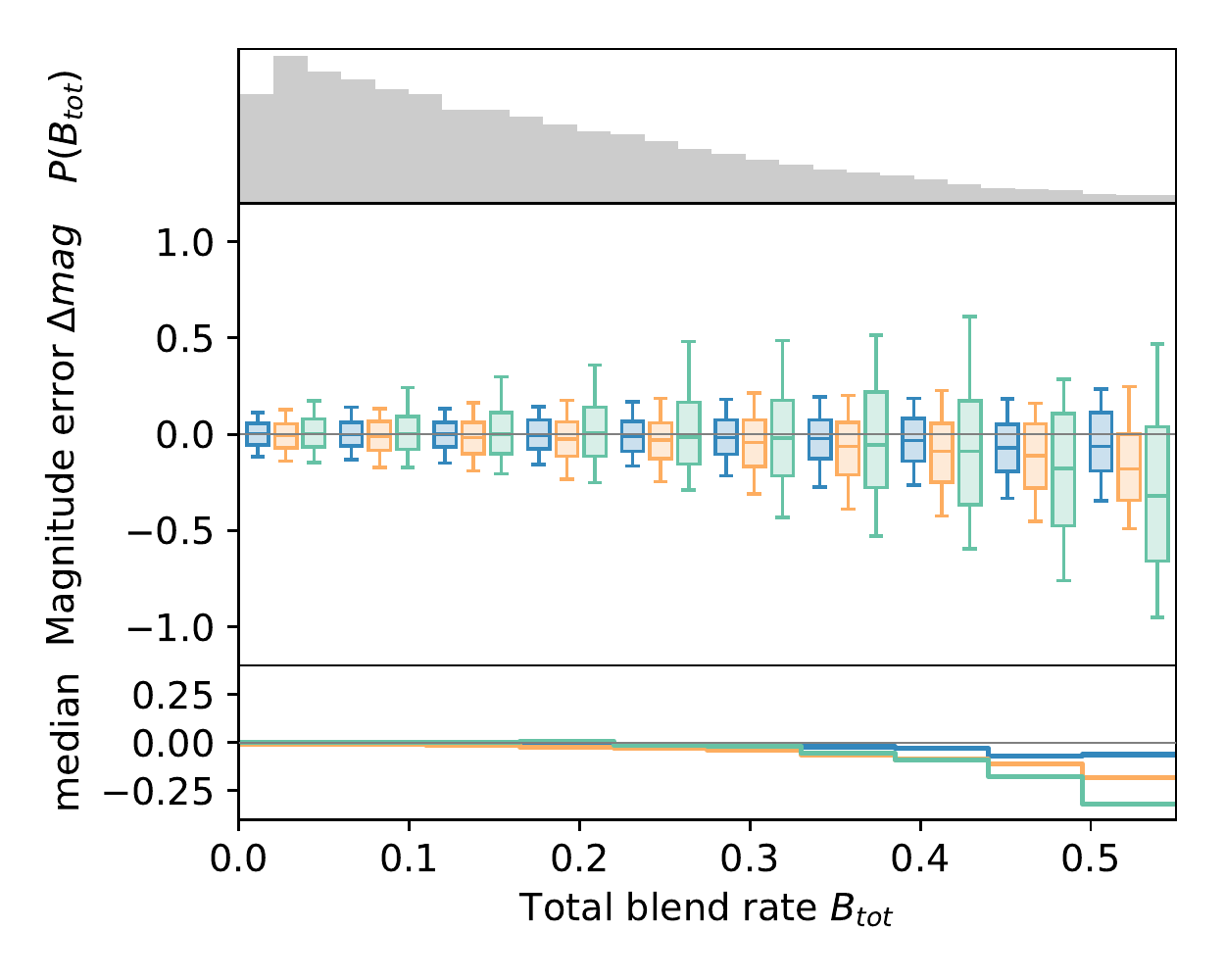}
        \\
        \text{b) Error in magnitude}
    \end{minipage}
    \caption{Comparison of results for LSST deblenders trained with three different centring configurations: the first one is perfectly centred on the stamp, the second one has its centre in a square of size \SI{0.2}{\arcsecond} around the stamp centre, and the third one has its centre computed by a simple peak detection algorithm (from the \texttt{photutils} library) on blended galaxies images (see \cref{sec:decentring_data}). The poor performance of the peak finder hamper those of the deblender network, which, in contrast, seems robust to decentring only due to pixelisation.}
    \label{fig:comparison_centreing}
\end{figure*}

In this section, we study the impact of decentring on deblending performance. We compare results obtained in the previous section on perfectly centred images to the two decentring configurations discussed in \cref{sec:decentring_data}, \ie either a shift within a square of side \SI{0.2}{\arcsecond} around the centre or a shift applied after centroid location obtained with a basic peak detection algorithm. In this part, images are composed of the 6 LSST bandpass filters only. We first train VAE on isolated galaxies in a similar fashion to our fiducial case, except that they are now off-centre. We have obtained error distributions qualitatively and quantitatively very similar to those presented in \cref{sec:vae_results} and \cref{fig:results_vae}, with median errors and distribution width only degraded by a few percents in both cases. We therefore directly present results on deblending performance.

\Cref{fig:comparison_centreing} shows the impact of decentring on the distributions of ellipticity and magnitude errors as a function of SNR and total blend rate $B_{\rm tot}$. In the first decentring configuration, the impact is relatively limited. Quantitatively, the ellipticity and magnitude median errors increase by a factor 1.4 to 3.5 and 1.4 to 2.8 respectively, but mostly remaining within $\pm\num{0.02}$ and $\pm\num{0.05}$ (except for the two last bins in total blend rate and first bin in signal-to-noise-ratio) and the spread of error distributions increase by 12\% and 22\% on average respectively for ellipticity and magnitude errors. In the second decentring configuration, we observe a general increase of the width of error distributions. While it performs similarly at $S/N\gtrsim60$ or blend rates below 25\%, we observe significantly degraded performance at low SNR or high blend rates. Overall, the median errors on the whole test sample are increased by a factor of 2 to 6.5 and 1.6 to 3.2 respectively for ellipticity and magnitude. The spread is then increased by 39\% for ellipticity and 53\% for magnitude in average.

\begin{figure}
    \centering
    \includegraphics[width=\columnwidth]{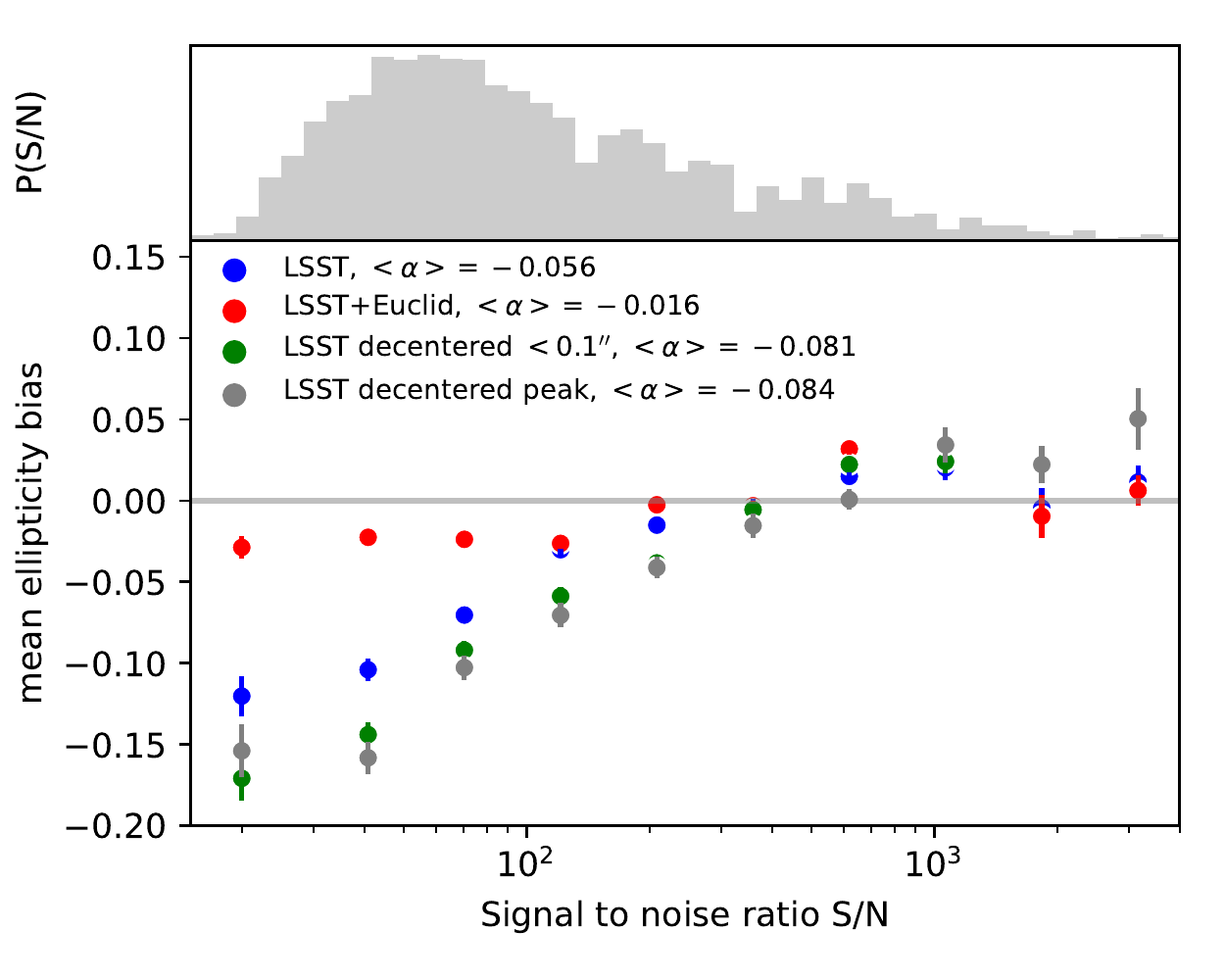}
    \caption{Multiplicative ellipticity bias (averaged over the two ellipticities components) between target and deblenders' output images, as a function of SNR in the four tested configurations: LSST and LSST+Euclid filters on perfectly centred target galaxies and two decentring configurations (simulating pixelisation and a simple peak finder algorithm, see \cref{sec:decentring_data}).}
    \label{fig:shear_bias}
\end{figure}

We further evaluate the impact of decentring by computing mean multiplicative ellipticity biases $\alpha_{i}$ for $i=1,2$, defined in \cref{eq:ellipticity_shear_bias}, by fitting a linear regression between output and input ellipticities at zero shear. We compute these biases introduced by our deblender as a function of SNR, which is shown on \cref{fig:shear_bias} for the four tested configurations. We find negative biases that decrease in amplitude for higher SNR and increase as we probe more complex decentring procedures. In the case of perfectly centred galaxies, we obtain an overall bias of \num{-0.056} for LSST alone, which is significantly reduced to \num{-0.016} for LSST+Euclid. On the decentred configurations, we measure biases of \num{-0.081} and \num{-0.084}.

This confirms that the deblender network can exploit colour information from the six LSST bands to perform deblending to a reasonable accuracy. In addition,
it can benefit from additional information provided by Euclid filters to improve shape reconstruction.
Finally, we have assessed the robustness of our method to decentring, whether it is due to pixelisation or to the accuracy of the peak detection algorithm.
These results thus emphasise the importance of an accurate detection pipeline for our method (as for any deblending technique) to work properly. In particular, the second case, which is strongly impacted by unrecognised blends, provides conservative performance tests. Our measurements might reveal the poor quality of the detection algorithm in case of blends rather than the actual performance of our deblender, as discussed further in \cref{sec:discussion_decentring}.

\section{Discussion}
\label{sec:discussion}

In this section, we further discuss some of our assumptions, in particular decentring, and their potential impacts on our results. We provide some insight into assets and potential caveats of the current implementation and propose avenues to solve some of our limitations --some of which will be challenging for any machine-learning-based deblending methods. We also discuss promising tests performed on real images from the COSMOS catalogue. 

\subsection{Detection and decentring}
\label{sec:discussion_decentring}

As mentioned in \cref{sec:decentring_results}, deblending is in practice closely tied to the detection pipeline. Indeed, blending affects both the pipeline's ability to detect multiple components and to locate centroids. In severe cases, two (or more) blended objects may be detected as a single one (\textit{unrecognised blends}), in which case the centroid is likely off by a significant amount to any true centroids \citep[see][for a study using the HST and the Subaru telescope]{2016ApJ...816...11D}. In this paper, we have proposed a method for deblending, which like most deblenders, requires some information about centroids. In our case, we only need the centroid of the galaxy to be deblended --but not the number of objects or their centres.

However, centroids necessarily include some error. We therefore studied the impact of decentring by considering two cases: an optimistic case including pixelisation where errors have a uniform distribution of about one LSST pixel, and a more conservative case based on a simple, single-band peak detector used to both detect the brightest galaxy within a blended scene and locate its centroid. 
The first configuration leads to only slightly degraded errors, demonstrating that our method can deal with sub-pixel decentring.
The second configuration results in more significant impacts on the spread of reconstruction errors, especially for critically blended scenes and at low SNR. However, several comments can be made.
First, we underline that we considered a simple peak detection algorithm and one might naturally assume that a more elaborate detection algorithm making use of multiple bands (and/or instruments) could perform better in blended scenes, leading to improved deblending performance.
Second, we also used this algorithm to generate the training samples (to the exception of rejecting galaxies too close to each other), which is not optimal as difficult cases may hamper optimisation. In particular, unrecognised blends often entail large decentring, yet, they form the majority of the training sample due to the poor performance of the detection algorithm. At last, another caveat is that the network's objective is also modified to deblend the brightest detected centroid galaxy in the $r$-bandpass filter instead of the smallest magnitude one, which we found to differ in 20\% of the training sample. Taking these different considerations into account, we suspect that the performance of our deblending method are indirectly degraded by those of the detection algorithm. We expect that they would be improved by an algorithm suited to deal with blended scenes, potentially one based on neural networks as well. We therefore consider those results to be somewhat of a lower bound on the performance of our method.

Furthermore, we envisage the optimal way to isolate every galaxy within a blended scene with our method would involve an iterative procedure of peak detection and deblending. With this procedure, one does not need to make assumptions about the number of objects in the first pass, unlike most other deblending algorithms, such as \scarlet, for instance. This process might be helpful in unrecognised blend cases, where faint, undetected galaxies could be identified after a few iterations.

Finally, we have made the assumption that all galaxies in our sample have a counterpart in both LSST and Euclid, regardless of their magnitude or other properties. For galaxies between the Euclid and LSST magnitude limits, Euclid images are noise-dominated and therefore do not carry much information. However, Euclid images could be used in blended scenes to obtain centroids for galaxies above the Euclid magnitude threshold and facilitate deblending, thus making the most of the high resolution of the VIS instrument for a joint, pixel-level analysis.

\subsection{Challenges with real data}
\label{sec:real data}
In this work, we have used simulated images allowing us to compare our results to a ground truth, a common practice to calibrate and validate algorithms in weak lensing studies \citep[see, for instance,][]{2018MNRAS.481.3170M}. However, the driving motivation in employing machine-learning techniques in general is to use real data to build more realistic models that reproduce the diversity of galaxy morphologies (including colour gradients) and observational systematic effects.

\begin{figure}
    \centering
    \includegraphics[width=\columnwidth]{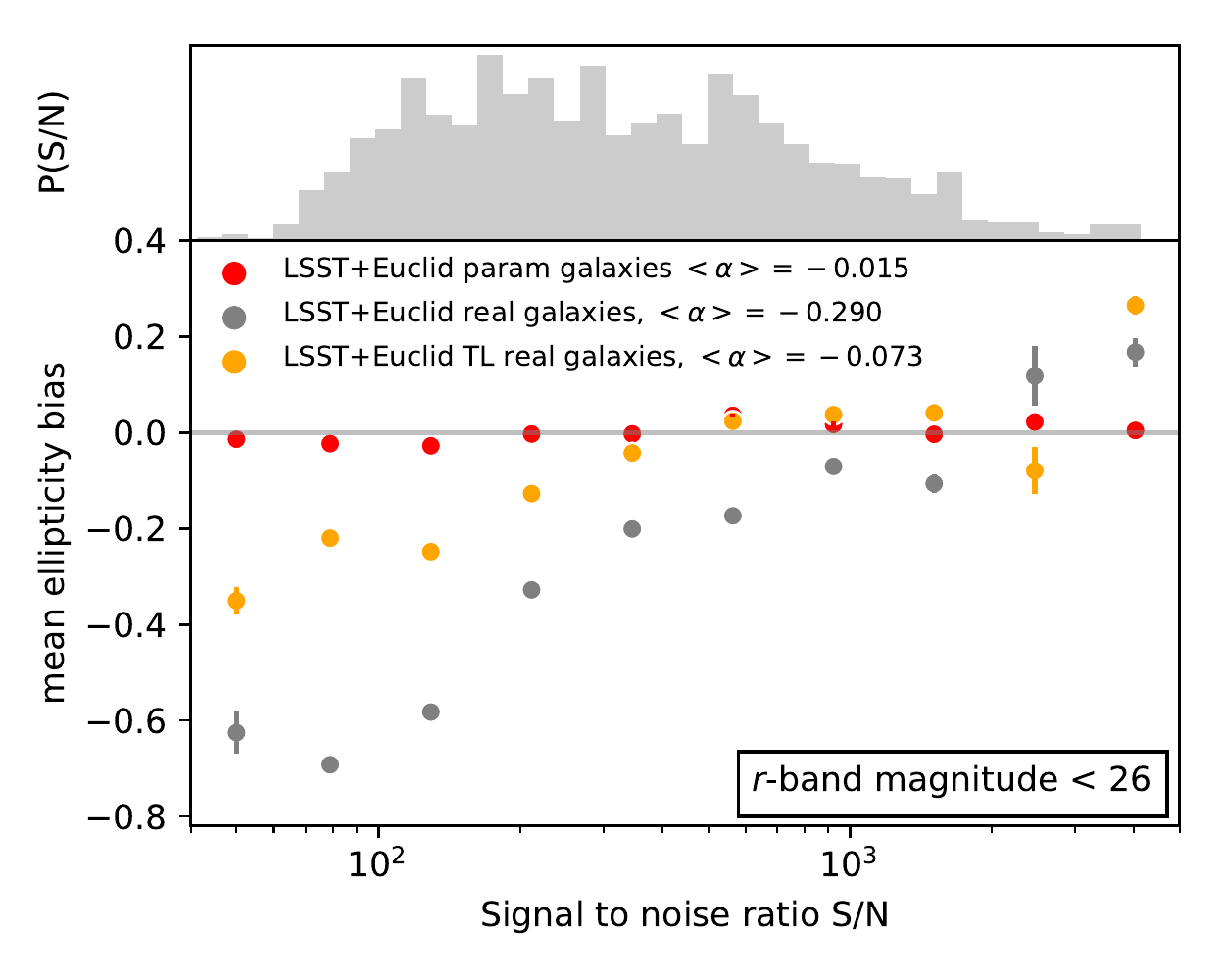}
    \caption{Multiplicative ellipticity bias (averaged over the two ellipticity components) between target and deblenders' output images, as a function of SNR in three tested configurations: LSST+Euclid filters trained and tested on parametric images (in red), LSST+Euclid filters trained on parametric images and tested on real images (in grey), and LSST+Euclid filters first trained on parametric images, then on real images (applying \textit{transfer learning}) and tested on real images (in orange). This figure is realised for galaxies with a minimal magnitude of 26 in the $r$-band filter.}
    \label{fig:shear_bias_transfer_learning}
\end{figure}

We have therefore tested our deblender network on more realistic images. In concrete terms, we have generated images with \galsim from the COSMOS catalogue, but this time using the real images of galaxies to capture more complex morphologies, which we then rescaled in different bands to reproduce the flux of their parametric counterparts. We input those to trained deblenders  and obtained promising results (see \cref{app:Deblending_real_images}). We also trained deblender networks on a sample of "real" images, using networks that were pretrained on parametric images, a technique known as \textit{transfer learning} \citep[see, for instance,][for an application in galaxy classification]{2019MNRAS.484...93D}. This significantly improves reconstruction, decreasing the multiplicative ellipticity bias by a factor of about 2 on the entire test sample ($r$-band magnitude < 27.5) and by a factor of about 4 when considering only galaxies with $r$-band magnitude < 26 (see \cref{fig:shear_bias_transfer_learning}).
However, a major obstacle is that HST images have a limiting depth of 25.2 ($i$-band), making it difficult to produce effectively noiseless target images, and they present correlated pixel noise that cannot be straightforwardly corrected for with \galsim, making adaptation difficult. Moreover, this sample still contains clear blends and residuals of image processing, despite exclusion cuts already available in \galsim (see, for instance, the last two rows of \cref{fig:deblender_real_images}).

This test suggests that assembling a clean and complete training sample from observations is challenging, as deep field data, which have higher SNR, also present more blending in the first place. Moreover, selecting isolated objects could introduce selection biases, for instance isolated galaxies are harder to find in high density regions such as galaxy clusters. Nevertheless, a plausible avenue will be to first assemble a small training sample of images of confirmed isolated galaxies, possibly through visual inspection or another network specialised in detecting blends. Then, this could be combined with a larger sample of realistic simulated images, in order to properly apply transfer learning. Large, realistic image simulations including all known instrumental and observational effects (\eg quantum efficiency of CCDs, electronic read-noise, brighter-fatter effect, persistence of IR detectors, asteroids, cosmic rays) are already being produced for upcoming surveys (as part of the Data Challenge 2 and 3 simulations for LSST Dark Energy Science Collaboration, see \cite{2020arXiv200100941S} and  \cite{2019ApJS..245...26K}), from which it will be possible to compile high-quality training samples for blending-related algorithms. 

Finally, we note that another possible avenue to test the deblender on more realistic scenes consists in injecting simulated galaxy images in real ones. This method does not solve the problem of training a generative model on real galaxies, but it would assess the ability of the network to recover the learned parametric model from images including other real galaxies, but also stars and any observational defects.

\subsection{Towards a Bayesian weak-lensing pipeline}
\label{sec:towards_bayesian}

As explained in \cref{sec:VAE}, the variational autoencoder output consists of a distribution over image space (usually taken as the product of independent distributions for each pixel in each band). The input should therefore be compared to samples drawn from the output distribution. However, as explained in \cref{sec:architecture_VAE}, we interpreted, as is common practise, the networks' output as an image. We therefore commit an abuse as the Bernoulli distribution does not correspond to our data which is distributed in the range [0,1].

The output of our method is probabilistic in the sense that we do sample latent variables according to the approximate posterior.
We show in \cref{app:bayesian_output} examples of such sampling which allow us to obtain a posterior distribution on the ellipticity and magnitude of a target, deblended galaxy.
Note, however, that the approximate posterior is computed from blended galaxy images while the decoder computes a posterior from a prior trained on isolated galaxy images. This difference breaks the Bayesian property of the loss function and consequently, authors warn that using the deblender's output to feed a Bayesian pipeline could lead to misleading results.

A solution to both aforementioned issues, which will be studied in subsequent work, is to directly generate multivariate posteriors on shape and flux parameters.

\section{Summary and Conclusions}
\label{sec:conclusion}

Blending of galaxies will have a major impact on weak lensing studies of future surveys, especially deep, ground-based ones such as LSST.
In this paper, we have investigated a new method based on deep, probabilistic neural networks, namely variational autoencoders (VAE), to learn a generative model of isolated galaxy images, that we then use as a prior to performing deblending itself. For this second step, we create another network, similar to the VAE, where the generative model is kept fixed. This amounts to perform deblending in a latent space where data is embedded in a probabilistic manner. Our study is meant as a proof of concept and used simulated images emulating observations from the LSST and the Euclid surveys. They are generated with the \galsim software and based on a public catalogue from observations of the COSMOS field by the Hubble Space Telescope.

We found that VAEs parametrised by convolutional neural networks (CNN) provide powerful models able to learn features of galaxy images from a noisy sample with sufficient accuracy to recover ellipticities and magnitudes. Then, the second network learns to deblend input images by recovering posterior parameters, that is, the distribution of latent variables encoding the target galaxy.
We evaluated the performance of the deblender neural network in terms of statistical reconstruction errors on ellipticities and magnitudes, and demonstrated that the network was able to isolate galaxies and recover their properties with low biases.
In particular, we found that the deblender network trained on LSST $ugrizy$ images was able to reach a median error on shapes contained within $\pm\num{0.01}$ and on $r$-band magnitudes within $\pm\num{0.05}$, stable across signal-to-noise ratios spanning the range 10 to 3000. We were able to further decrease the shape median error by 8 to 47\% and of the widths of error distributions by about 33\% when using 10-band images using all of LSST and Euclid filters (including visible and near-infrared filters). We also measured multiplicative ellipticity biases of respectively 5.6\% and 1.6\% for LSST and LSST+Euclid images, and uncalibrated multiplicative shear bias of the order of 1 to 4\% for both configuration, averaged over the full test sample of blended images.
Therefore, our method makes efficient use of the six LSST bands that provide crucial information to distinguish multiple sources and perform deblending.
It also naturally integrates information from multiple instruments to learn tighter posteriors.

We then studied the impact of decentring due to pixelisation and errors in peak detection and centroid measurement. We considered two cases, one optimistic, addressing mostly pixelisation and supposing accurate centroid localisation, and a more conservative case where we applied a simple peak finder on highly blended scenes. Even in the conservative case, our method produces a mean ellipticity bias on the measured ellipticities of 8.4\%, improving with signal-to-noise ratio, which is slightly reduced to 8.1\% in the optimistic case. The latter yields accurate ellipticities $|e|$ and magnitude with median errors below \num{0.03} and \num{0.2} (in absolute value), stable with signal-to-noise ratio and blending rates. Our method is consequently robust to pixelisation-related and/or modest decentring. It degrades, particularly for the spread of the distributions, with larger decentring errors caused by the associated peak detection algorithm, but would likely improve along with a more accurate detection pipeline suited to blended images. A potential avenue to measure and mitigate those biases, which we explore in \cref{sec:vae_results}, is to apply a technique resembling the metacalibration \citep{2017arXiv170202600H,2017ApJ...841...24S} and metadetection algorithms \citep{2019arXiv191102505S} once training is complete.
Moreover, the envisaged procedure to isolate every galaxy within a blended scene would involve iterations of peak detection and deblending. This procedure, contrary to most other deblending algorithms, presents the major advantage of not needing to make assumptions about the number of objects in the first pass. We reserve implementation and testing of this procedure for future work.

The next step is to adapt our method to real images. We have outlined challenges in assembling a clean training sample without unrecognised or unprocessed blends. We nonetheless conducted preliminary tests by applying transfer learning techniques with COSMOS data, as described in \cref{sec:real data}, and obtained promising results. We therefore strongly recommend to explore this direction for learning-based deblending algorithms. We also note that variational autoencoders have been used in the literature as denoisers, a feature of potential interest for weak lensing measurement pipelines.

Another avenue, that we reserve for future work, is to replace the decoder by a simpler network that would parametrise the joint likelihood of the ellipticities and magnitude, and potentially other quantities such as redshift. In order to efficiently estimate the posterior on those parameters, one would consecutively sample latent variables and the likelihood, thus providing a weak lensing analysis pipeline with a Bayesian input.

Finally, we have shown that our deblender networks are able to make the most of multi-bands and multi-instruments images to retrieve tighter posterior and likelihood distributions. The higher resolution of the Euclid VIS instrument provides additional information to LSST bandpass filters, yielding significant reductions of errors and biases on galaxy shapes even at high blend rates and low SNR.
We hope our results can encourage collaborations between future photometric surveys and the development of joint-pixel analysis and simulation tools.

\section*{Acknowledgements}
This paper has undergone internal review in the LSST Dark Energy Science Collaboration. The authors would like to kindly thank the internal reviewers: David Kirkby, François Lanusse and Fred Moolekamp. Their comments, feedback and suggestions helped us greatly improve the paper. We also would like to thank Gary Bernstein, Alexandre Boucaud, Patricia Burchat, Catherine Heymans, Bhuvnesh Jain, Mike Jarvis, Rémy Joseph, Sowmya Kamath, Robert Lupton, Peter Melchior, Jason Rhodes and Robert Schuhmann for comments and discussions throughout this work. We finally thank the anonymous reviewer for useful suggestions and comments.

Authors contributions are listed here.
B.A. and C.D. co-led the project, wrote the code, trained the neural networks, performed the analysis and wrote the paper (initial draft and editing after internal reviews). E.A. and C.D. designed the project. E.A and C.R. supervised the project.

The authors benefited of Cloud Credits for Research from Amazon Web Services. We also would like to acknowledge the support from the Paris Center for Cosmological Physics.

The DESC acknowledges ongoing support from the Institut National de Physique Nucl\'eaire et de Physique des Particules in France; the Science \& Technology Facilities Council in the United Kingdom; and the Department of Energy, the National Science Foundation, and the LSST Corporation in the United States.  DESC uses resources of the IN2P3 Computing Center (CC-IN2P3--Lyon/Villeurbanne - France) funded by the Centre National de la Recherche Scientifique; the National Energy Research Scientific Computing Center, a DOE Office of Science User Facility supported by the Office of Science of the U.S.\ Department of Energy under Contract No.\ DE-AC02-05CH11231; STFC DiRAC HPC Facilities, funded by UK BIS National E-infrastructure capital grants; and the UK particle physics grid, supported by the GridPP Collaboration. This work was performed in part under DOE Contract DE-AC02-76SF00515.

C.D. acknowledges partial support from NASA ROSES grant 12-EUCLID12-0004.

We used various open-source libraries including \galsim, \texttt{TensorFlow}, \texttt{Keras}, \texttt{photutils}, \texttt{astropy} and \texttt{tkiz}\footnote{\url{https://github.com/jettan/tikz_cnn}}. This work was based on the COSMOS catalogue packaged with \galsim and built from observations of the Hubble Space Telescope.

\section{Data availability}
The HST COSMOS catalogue used in this article is available at \url{https://zenodo.org/record/3242143}.





\bibliographystyle{mnras}
\bibliography{biblio.bib} 

\appendix

\section{Examples of deblended scenes}
\label{app:Deblending_images}

\begin{figure*}
    \centering
    \includegraphics[trim=24cm 13.2cm 6cm 6cm,clip, width=0.92\textwidth]{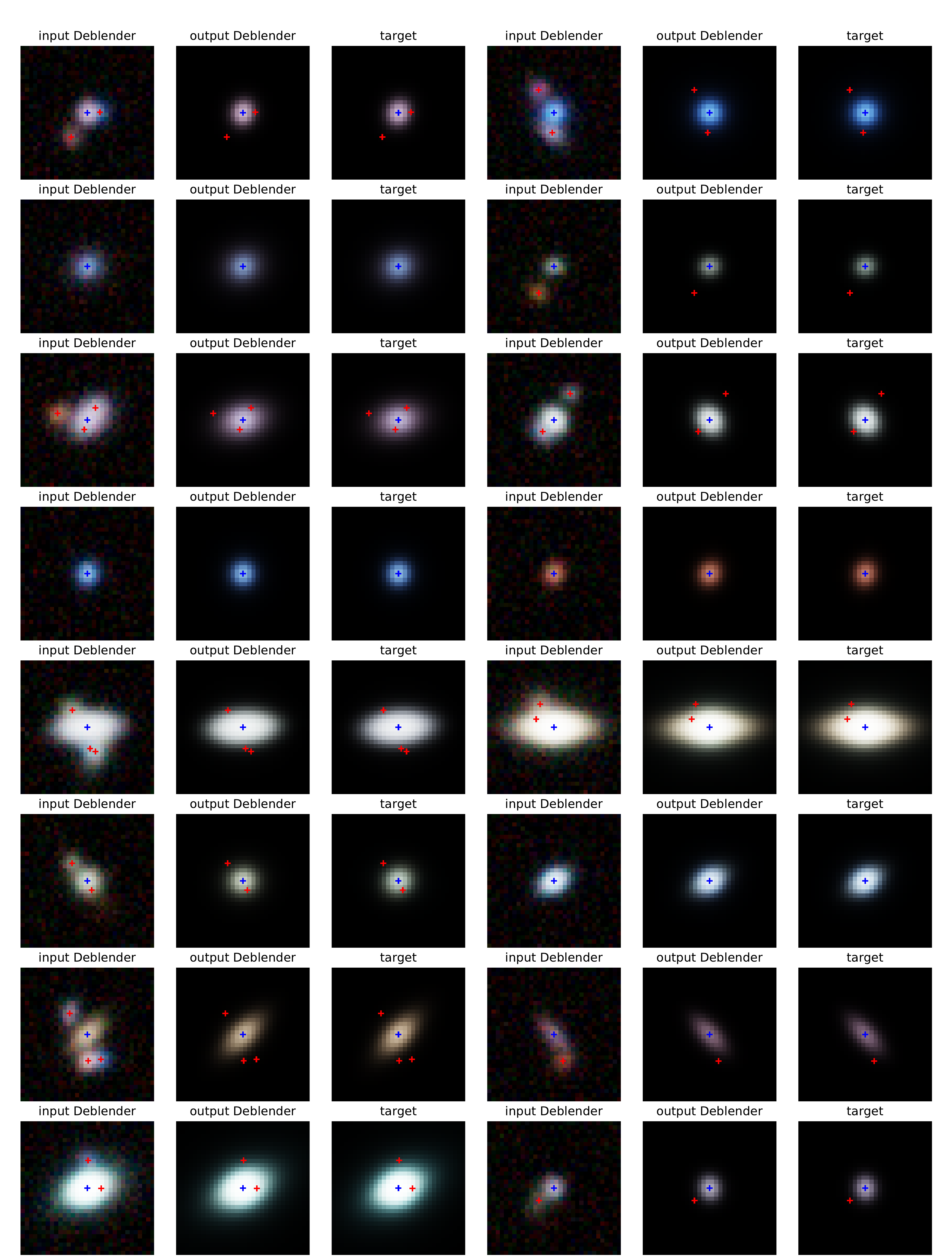}
    \caption{Random sample of normalised images processed by the deblender trained on both LSST and Euclid bandpass filters, where the target galaxy is perfectly centred on the stamp. We show two examples per row, with the first image showing the noisy input image, the second the output of the deblender and the third the noiseless target. These images are all shown in the $gri$ bands using the normalisation defined in \cref{sec:data_sample}. On each image, the blue cross indicates the centroid of the target galaxy and the red ones those of other galaxies in the scene. Images have been cropped to 50\% for improved visualisation.}
    \label{fig:deblender_images}
\end{figure*}

\Cref{fig:deblender_images} shows a random selection of deblended scenes.

\section{Deblending real images}
\label{app:Deblending_real_images}

\begin{figure*}
    \begin{minipage}[b]{.95\textwidth}
        \centering
        \includegraphics[trim={0cm 2.1cm 0 1.1cm}, clip, width=0.8\linewidth]{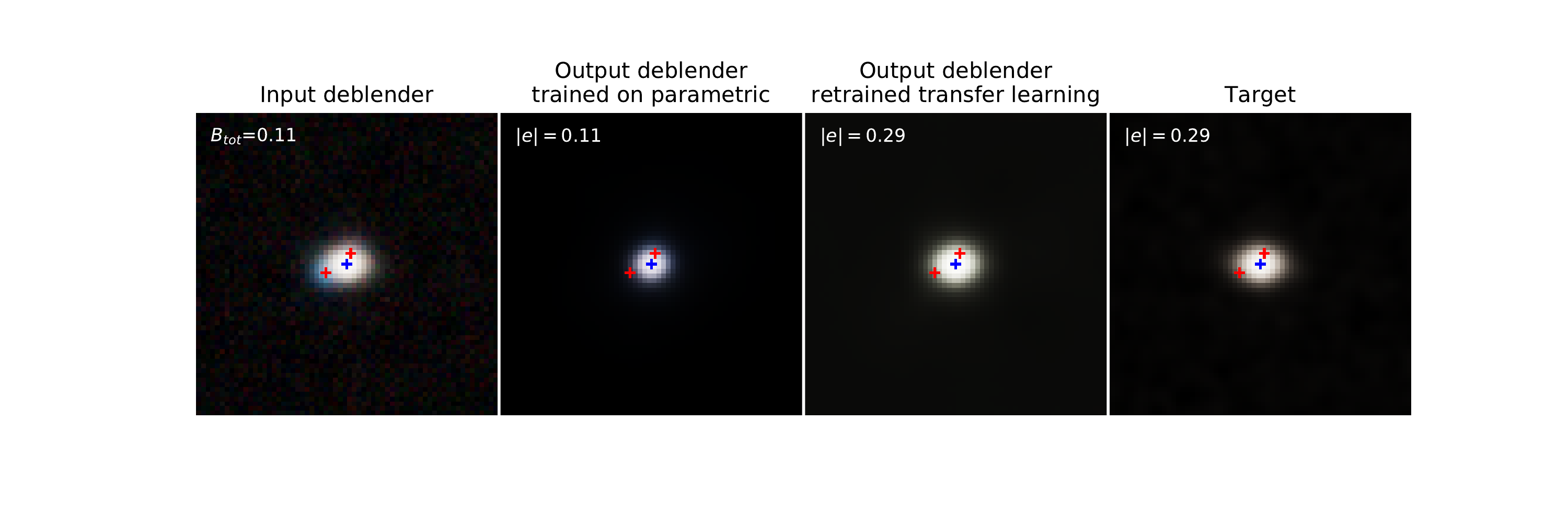}
        \includegraphics[trim={0cm 2.1cm 0 1.1cm}, clip, width=0.8\linewidth]{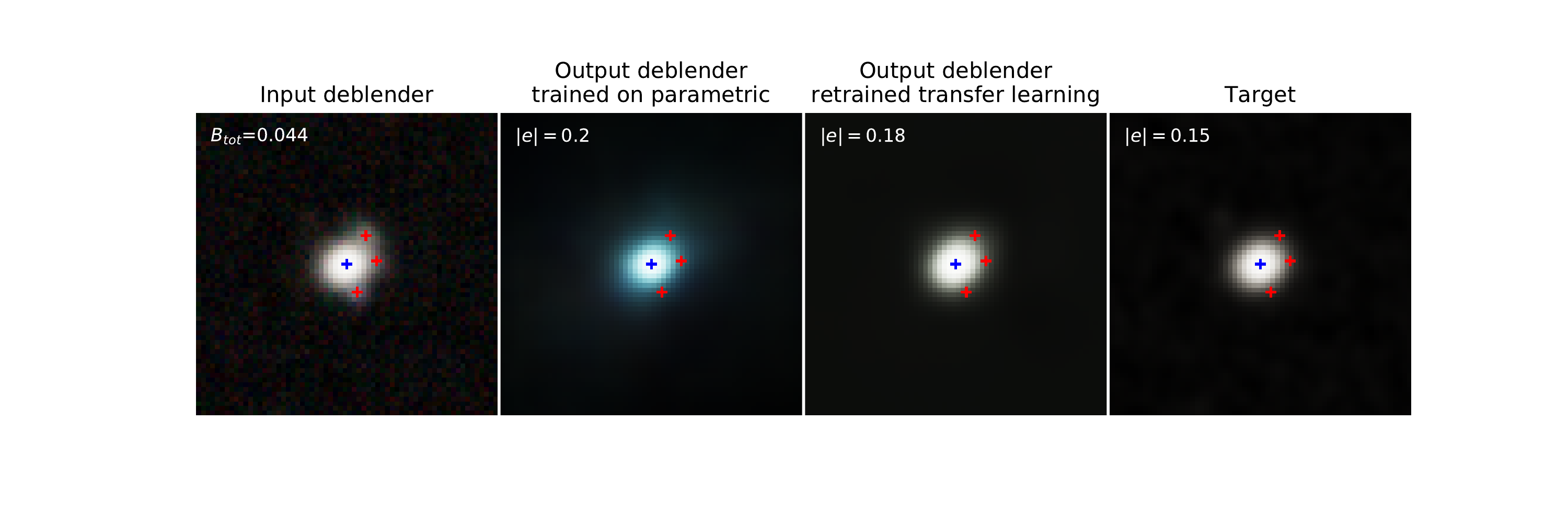}
        \includegraphics[trim={0cm 2.1cm 0 1.1cm}, clip, width=0.8\linewidth]{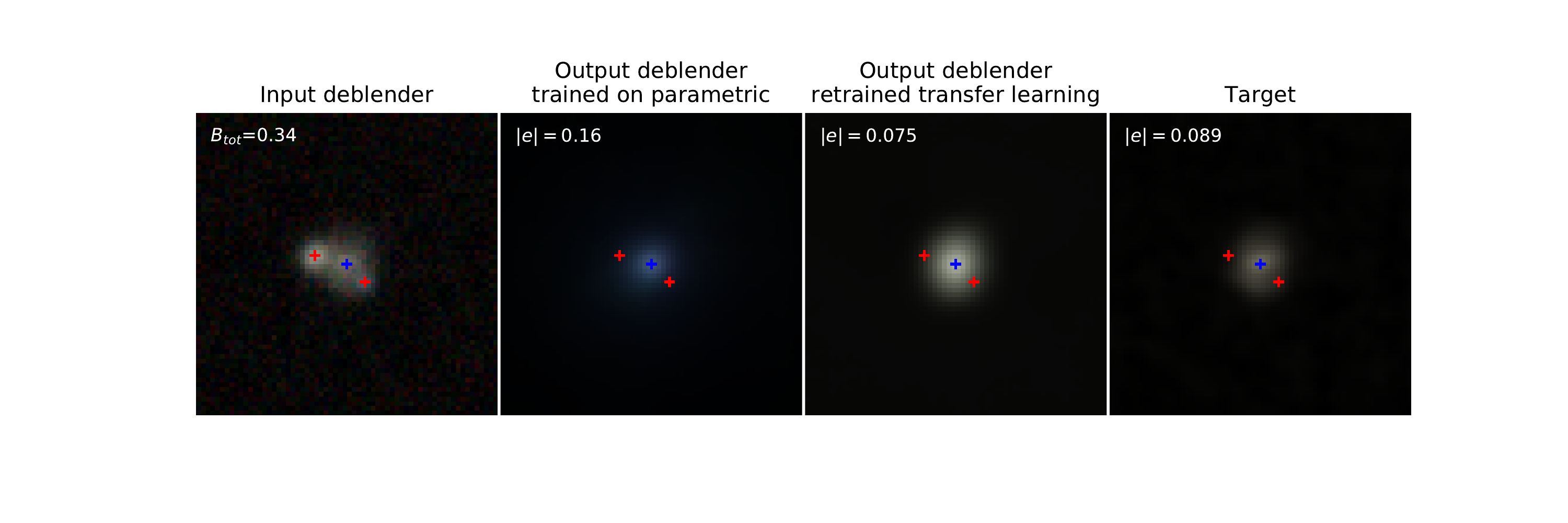}
        \includegraphics[trim={0cm 2.1cm 0 1.1cm}, clip, width=0.8\linewidth]{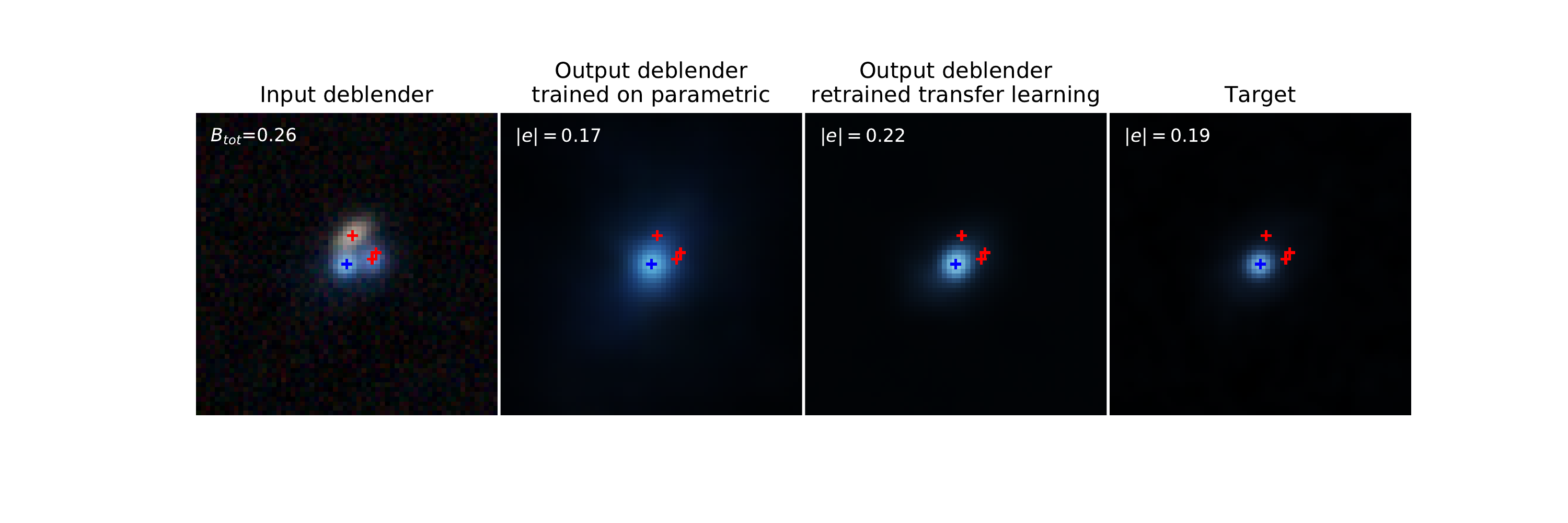}
        \includegraphics[trim={0cm 2.1cm 0 1.1cm}, clip, width=0.8\linewidth]{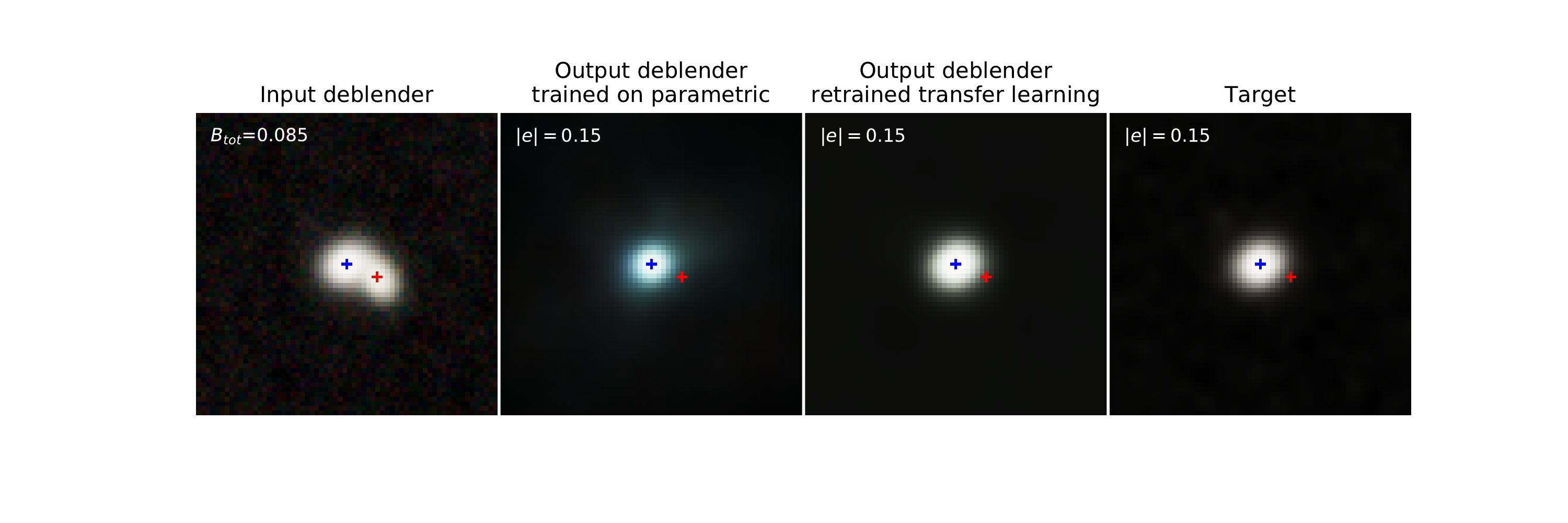}
        \includegraphics[trim={0cm 2.1cm 0 1.1cm}, clip, width=0.8\linewidth]{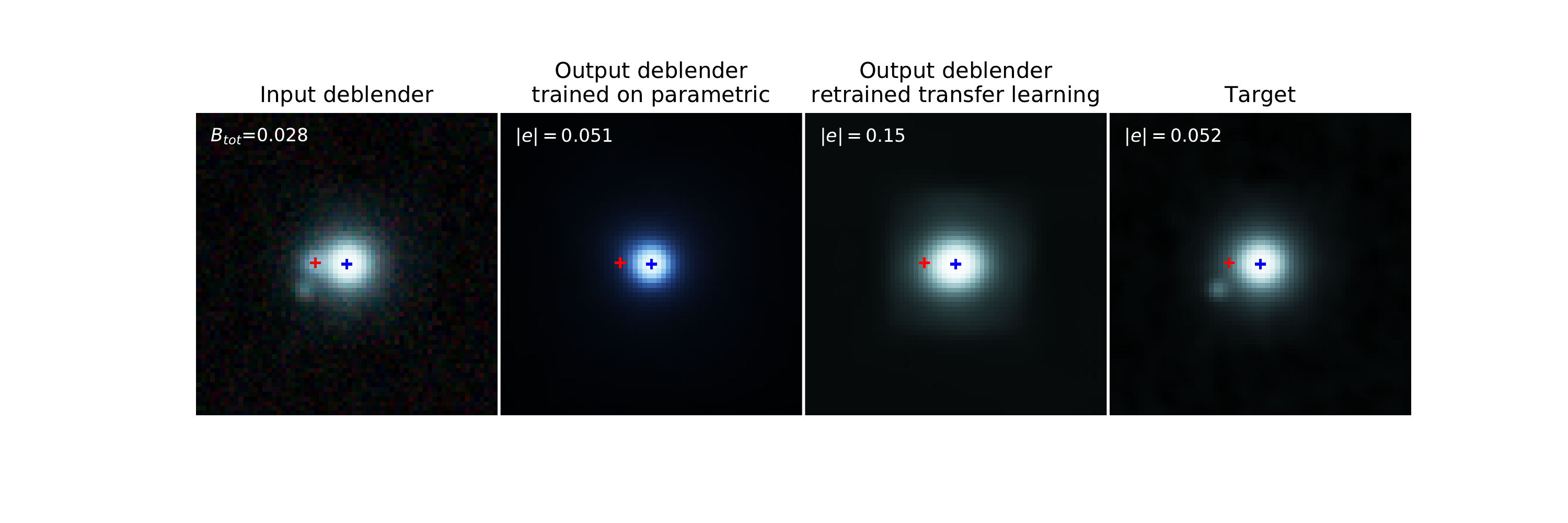}
        \includegraphics[trim={0cm 2.1cm 0 1.1cm}, clip, width=0.8\linewidth]{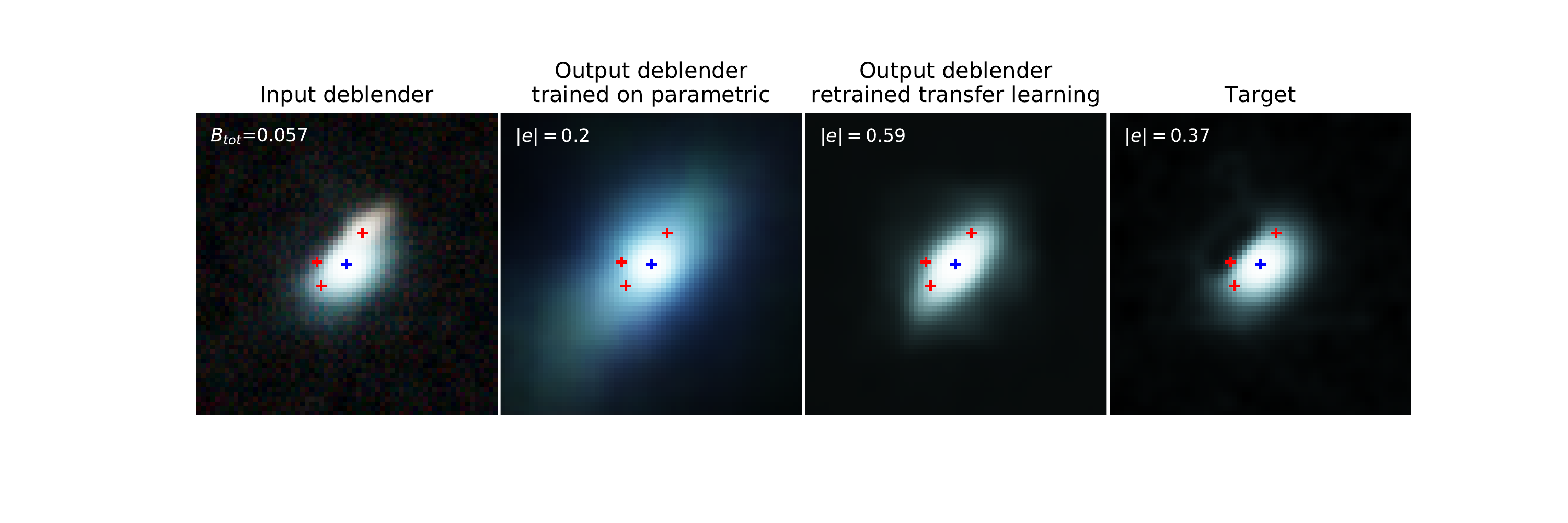}
    \end{minipage}
    \caption{Sample of images processed with the LSST+Euclid deblender networks, before and after applying transfer learning. The first column shows the noisy input image and the fourth one the target galaxy image. As mentioned in \cref{sec:real data} correlated noise and residuals of images processing appear in target images. The middle columns show the output of the network when only trained on simulated images (second from left) and after retraining on a sample including 20\% of real images (third from left).}
    \label{fig:deblender_real_images}
\end{figure*}

\Cref{fig:deblender_real_images} shows a random selection of deblended images using real galaxy images. Results are shown for a deblender trained only on parametric models and for one trained with transfer learning, i.e. with artificially blended images composed of real galaxy after a training on parametric models.

\section{Probabilistic output}
\label{app:bayesian_output}

\begin{figure*}
    \begin{center}
	    \includegraphics[width=0.7\textwidth]{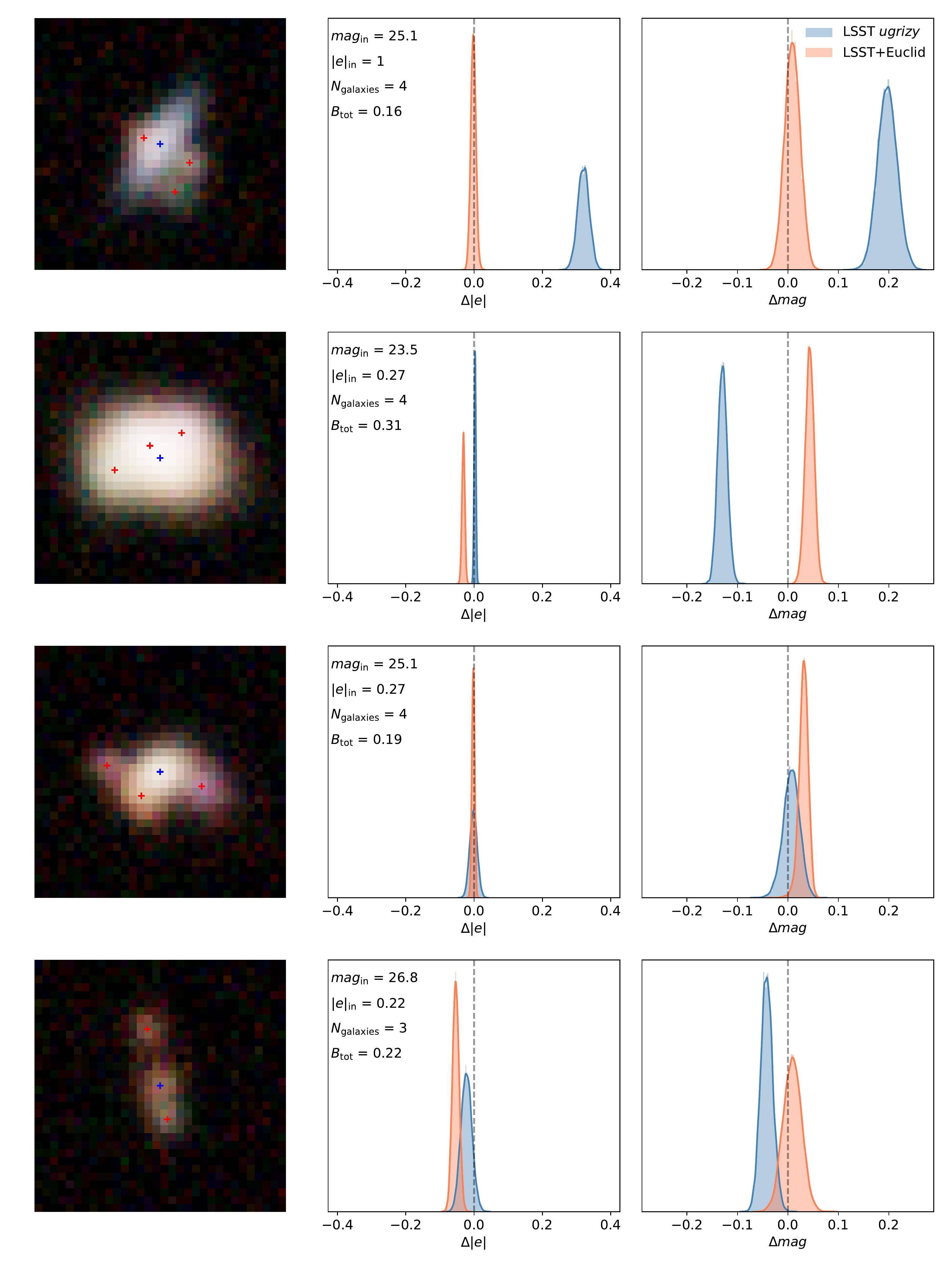}
        \caption{Distributions of ellipticity and magnitude errors for individual images (shown on the left) passed \num{10000} times in the LSST and LSST+Euclid deblenders. For each image, the encoder parametrises the approximate posterior from which latent variables are sampled.}
        \label{fig:VAE_proba}
    \end{center}
\end{figure*}

For all figures in \cref{sec:results}, we pass each image in the test sample only once through the networks and use the output to estimate distributions of errors on ellipticities and magnitudes. However, as explained in \cref{sec:towards_bayesian}, each passage requires sampling latent variables, making the output probabilistic. As an illustration, we consider random test images that we each pass \num{10000} times in the deblender network in order to obtain the corresponding distributions of ellipticity and magnitude errors, as shown in \cref{fig:VAE_proba}.

\bsp	
\label{lastpage}
\end{document}